\documentclass[10pt,journal,final,letterpaper,twocolumn,twoside]{IEEEtran} 

\usepackage[dvips]{epsfig}
\usepackage[dvips]{graphicx}
\usepackage{amsmath,amsfonts,bm,amssymb,psfrag,ifthen,color,subfigure,amsthm}
\usepackage{algpseudocode}
\usepackage{algorithm}
\usepackage{algorithmicx}
\usepackage{setspace}
\usepackage{cite}
\usepackage{url}
\usepackage{longtable}
\usepackage{stfloats}
\usepackage{graphics,booktabs,color,subfigure}
\usepackage{float}
\usepackage{diagbox}
\usepackage{multirow}

\allowdisplaybreaks[4]

\newcounter{mytempeqncnt}

\begin{document}
\title{ Optimal  Discrete Constellation Inputs for Aggregated LiFi-WiFi   Networks}
 \author{Shuai~Ma, Fan Zhang, Songtao Lu, Hang Li,~\IEEEmembership{Member,~IEEE},  Ruixin Yang, Sihua Shao,  Jiaheng Wang,~\IEEEmembership{Senior Member,~IEEE},  Shiyin Li
\thanks{ S. Ma, F. Zhang, R. Yang and S. Li   are with the School of Information and Control   Engineering, China
University of Mining and Technology, Xuzhou 221116,
China. (e-mail:  \{mashuai001;zhangfan; ray.young; lishiyin\}@cumt.edu.cn).}
\thanks{S. Lu  is with the IBM Thomas J. Watson Research Center, Yorktown
Heights, NY 10598 USA (e-mail: songtao@ibm.com).}
\thanks{H. Li is with the Shenzhen Research Institute of Big Data, Shenzhen 518172, Guangdong, China. (email: hangdavidli@163.com).}
\thanks{S. Shao is with the Department of Electrical Engineering, New Mexico Tech, Socorro, NM 87801 USA. (email: sihua.shao@nmt.edu).}
\thanks{J. Wang is with the National Mobile Communications Research Laboratory, Southeast University, Nanjing 210096, China,
and also with the Purple Mountain Laboratories, Nanjing 211111, China.
(e-mail: jhwang@seu.edu.cn).}

}

\maketitle
\begin{abstract}

In this paper, we investigate the performance of a practical aggregated LiFi-WiFi system with the discrete constellation inputs
from a practical view.
 We  derive the achievable rate expressions of the aggregated LiFi-WiFi system for the first time. Then, we study the rate maximization problem via optimizing the constellation distribution and power allocation jointly. Specifically, a multilevel mercy-filling power allocation scheme is proposed by exploiting the relationship between the mutual information and minimum mean-squared error (MMSE) of discrete inputs. Meanwhile, an inexact gradient descent method is proposed for obtaining the optimal probability distributions. To strike a balance between the computational complexity and the transmission performance, we further develop a framework that maximizes the lower bound of the achievable rate  where the optimal power allocation can be obtained in closed forms and the constellation distributions  problem can be solved efficiently by Frank-Wolfe method. Extensive numerical results show that the optimized strategies are able to provide significant gains over   the state-of-the-art schemes in terms of the achievable rate.

\end{abstract}

\IEEEpeerreviewmaketitle

\section{Introduction}

\subsection{Motivation and Contributions}

The   increasing number of
Internet of Things (IoT) devices exerts   tremendous bandwidth burden continuously on the wireless networks.
{According to Ericsson' report \cite{Ericsson}, more than $80\%$ of the wireless data are generated in indoor environments. Visible light communication (VLC) or light fidelity (LiFi), with a vast license-free visible band in 400-790 THz, can    support both high speed data transmission and illumination simultaneously.}
    LiFi exploits   the   off-the-shelf    light emitting diodes (LEDs)   and photodiodes (PDs) as transceivers, which
  can be integrated into  IoT devices.
Although LiFi serves as a competitive candidate for the next generation wireless solution, its vulnerability to the blockage and small its signal coverage still impose many challenges on various indoor applications. Therefore, a more practical solution is to aggregate the LiFi and WiFi systems and exploit their unique advantages under certain conditions.

In this paper, we consider an   aggregated LiFi-WiFi system from a practical communication perspective. First of all, we derive the achievable rate expression of the system with the discrete constellation input signals, rather than the Gaussian inputs adopted in most of the existing works. Then, we further investigate the optimal input distribution and power allocation for the considered system. Our results provide a relatively practical design framework for the aggregated LiFi-WiFi communication system.  Specifically, the  main contributions of this work are given as follows:

\begin{itemize}

\item 
{ Generally, the  inputs of the practical communications systems   follow an finite-set discrete distribution rather than   Gaussian distributions.
    To obtain the performance description of the aggregated LiFi-WiFi system with  an arbitrary discrete distribution, we derive the  achievable rate expressions  of LiFi links and WiFi links, respectively. Comparing with the existing  rate expressions with equiprobable discrete constellation points, the derived  results are more general and practical.}
Given that such rate expression is not in closed-form, we further    derive both the
lower and the upper bounds. All these results can be used as the performance metric for the considered system.

\item  We  jointly
     optimize    the  discrete constellation input  distribution
and the power allocation   to maximize the derived achievable rate. To handle this non-convex problem,  we propose a multi-level mercy-filling method to obtain
 the suboptimal power allocation scheme,
 which    exploits the relationship between the mutual
information and the minimum mean square error (MMSE).
The optimal probability distributions of  the  discrete constellation are  calculated by the inexact gradient descent method.

\item {To reduce the computation complexity in the previous design problem, we further adopt    the derived lower bound as the performance metric. Specifically, we jointly optimize the discrete constellation input distribution and the power allocation to maximize the derived lower bound.
    To overcome the  difficulty of the nonconvex problem, we iteratively optimize  the power allocation and discrete  probability distribution, where the optimal power allocation scheme is derived  with  closed expression, and
    the discrete  probability distribution sub-problems is optimized via  the proposed Frank-Wolfe method.}

\end{itemize}

\subsection{Related Works and Organization}

 There are   two schemes for the LiFi-WiFi  transmission:     hard-switching (hybrid) \cite{Yunlu_2,Yunlu_1,Papanikolaou,Xuan,Jin,Yunlu,Wu,Basnayaka,Shao,Hammouda,Tabassum} and  aggregating \cite{Shao,Hammouda,Tabassum,Jingjing,Zhang,Ma_twc_202011}. The former  realizes the transmission via  LiFi link or  WiFi link,
while the later  uses both   links   simultaneously.
   In general, the  former scheme usually results in a lower spectral efficiency and may   cause frequent link switches.
The
LiFi-WiFi aggregated systems, on the contrary, can increase  the system data rate and  provide  reliable
communication.

The   LiFi-WiFi aggregated system receives great attentions in recent years. For example, by leveraging the bonding technique in the Linux operating
system, the authors in \cite{Shao}    prove that  the aggregated system outperforms the conventional WiFi.
Based on the cross-layer analysis of the physical and datalink
layers,
link selection approaches  were  proposed in \cite{Hammouda} to maximize the average data arrival rate  and minimize the non-asymptotic bounds on data
buffering delay.
In \cite{Tabassum}, the  coverage probability and
  rate were analyzed  for the
 aggregated RF/VLC networks.  It has been shown that selecting the correct intensity of OBSs (optical BSs) plays a crucial role, and the performance of the  aggregated scheme outperforms
all other schemes (RF-only, VLC-only, and opportunistic RF/VLC).
In \cite{Jingjing}, the    access point selection strategies  were optimized via the multi-armed bandit scheme.
In \cite{Zhang},   both subchannel allocation and
power control scheme were developed  for improving the energy   efficiency of
the aggregated
VLC/RF network.

In most of existing works, e.g. \cite{Shao,Yunlu,Wu,Basnayaka,Jingjing,Zhang,Hammouda,Tabassum,Ma_twc_202011}, the achievable rate of the considered system is derived based on the assumption that the input signal follows the Gaussian distribution.
However, Gaussian signals may not be an accurate representation of the inputs in practical communication systems.
{The inputs  of the practical RF-only system are always generated based on the discrete constellations,  such as pulse amplitude modulation (PAM), quadrature amplitude modulation (QAM), and phase shift keying (PSK), etc., rather than the Gaussian codebook. Thus, those resource  allocation strategies based on the Gaussian assumption would lead to serious performance loss. It has been shown that the mutual information maximization strategies can improve performance of the practical communications systems  \cite{Globally}. Meanwhile, it is   proved that the optimal inputs of the VLC-only system   follow a finite-set discrete distribution \cite{Ma_ISJ_2020}. Although   some existing works studied the achievable rate based on the   discrete constellation points with equal probability  \cite{cover,Gamal,Tse, Globally,Zengwcl2012,Shamai}, those results cannot be  directly extended to the LiFi-WiFi systems with arbitrary discrete inputs.}
Therefore, it is necessary to investigate the optimal transmission scheme with the discrete   inputs   for the aggregated LiFi-WiFi system.

The rest of this paper is organized as follows.   We provide  the model of the aggregated
 LiFi-WiFi   system  in Section II.
  The optimal power allocation and optimal probability distributions schemes  of the aggregated
LiFi-WiFi system are presented in Section III.
     In Section IV, we provide the solutions for the lower bound as the throughput metric.
        The simulation results are presented  in Section V.
  Finally, Section VI concludes
the paper.

\emph{Notations}:     ${\left(  \cdot  \right)^{\rm{T}}}$, $\left(\cdot\right)^*$,  $\left\|  \cdot  \right\|$, and ${\rm{Tr}}\left(  \cdot  \right)$ represent  the  transpose, conjugate, Frobenius norm,  and trace of a matrix  respectively.
 The Hadamard product of ${\bf{A}}$ and ${\bf{B}}$ is denoted as
 ${\bf{A}} \odot {\bf{B}}$.
$\mathcal{M} \triangleq \left\{ {1,2,...,M} \right\}$ and  $\mathcal{N} \triangleq \left\{ {1,2,...,N} \right\}$.

\section{System model}
\begin{figure}[htbp]
      \centering
	\includegraphics[height=3.5cm]{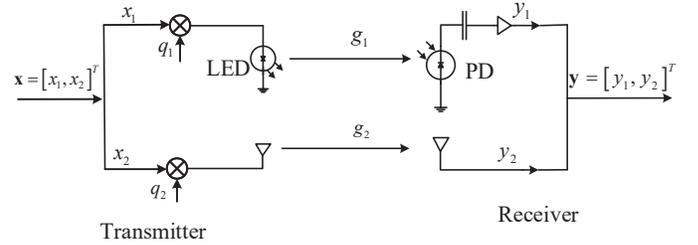}
      \vskip-0.2cm\centering
 \caption{ Schematic diagram of an aggregated LiFi-WiFi   system.}
  \label{LiFi_WiFi} 
\end{figure}
As illustrated in Fig. \ref{LiFi_WiFi},  we consider the downlink transmission of  an aggregated LiFi-WiFi  system, where the transmitter is equipped  with single LED and  single WiFi  antenna, and the receiver is equipped with single PD and single RF   antenna.
The transmitter   simultaneously  transmits information    via both the LiFi link and  the WiFi link,
where the bandwidths of LiFi link and  WiFi link are $B_1$Hz and $B_2$Hz, respectively.
 Let ${\bf{x}}\buildrel \Delta \over = \left[ {{x_{1}},{x_{2}}} \right]^T$ denote the transmitted signal vector, where ${x_1} \in {\mathbb{R}}$ and ${x_2}\in {\mathbb{C}}$ denote the {independently} transmitted signals  of the LiFi link and WiFi link, respectively.

In a   practical LiFi-WiFi communication
system,   the transmitted signals   are  distributed in
discrete constellation. Suppose that the LiFi link signal is sent via   $M$-pulse-amplitude modulation ($M$-PAM)      and the WiFi link signal is sent via $N$-quadrature amplitude
modulation ($N$-QAM). To be more specific,
 the signal $x_1$ is taken   from a  non-negative real  discrete
constellation set ${\Omega _1}$
  with cardinality $M$, which is given as
   \begin{align}
{\Omega _1} \buildrel \Delta \over =& \left\{ {{x_1}\left| \begin{array}{l}
\Pr \left( {{x_1} = {x_{1,k}}} \right) = {p_{1,k}},\\
0 \le {x_{1,k}} \le A,\\
\sum\limits_{k = 1}^M {{p_{1,k}}}  = 1,\\
\sum\limits_{k = 1}^M {{p_{1,k}}} {x_{1,k}}  \le \bar {\mu},\\
\sum\limits_{k = 1}^M {{p_{1,k}}} x_{1,k}^2 \le  P_{{\rm{e}},1},\\
 x_{1,k} \in {\mathbb{R}},k = 1,...,M,
\end{array} \right.} \right\},
\end{align}
where ${x_{1,k}}$ denotes the  constellation point,
 ${{p_{1,k}}} $   represents probability   that $x_1$   equals ${x_{1,k}}$, and parameters $A$,  $\bar {\mu}$,  and $  P_{{\rm{e}},1}$ denotes the peak optical
power,  maximum average optical
power,  and  maximum average electric   power of $x_1$, respectively.
  On the other hand,  the   WiFi signal   ${x_2}$   is taken   from a  complex discrete
constellation set  ${\Omega _2}$
  with cardinality $N$, which is given as
   \begin{align}
{\Omega _2} \buildrel \Delta \over = \left\{ x_1\left| \begin{array}{l}
\Pr \left( {{x_2} = {x_{2,l}}} \right) = {p_{2,l}},\\
\sum\limits_{l = 1}^N {{p_{2,l}}}  = 1,\\
\sum\limits_{l = 1}^N {{p_{2,l}}{{\left| {{x_{2,l}}} \right|}^2}}  \le {P_{{\rm{e}},2}},\\
{x_{2,l}} \in {\mathbb{C}},l = 1,...,N,
\end{array} \right. \right\}
\end{align}
where ${x_{2,l}}$ denotes the constellation point,
 ${{p_{2,l}}} $ denotes the probability that    the constellation ${x_{2,l}}$ is chosen, $ P_{{\rm{e}},2}$ denotes  the   maximum average electric   power of $x_2$, respectively.

 Let  ${q}_1  \in {\mathbb{R}}$ and ${{q}}_2 \in {\mathbb{C}}$ denote the power amplification factors for ${x_1}$ and ${x_2}$, respectively.
Here, ${q}_1 $ and ${{q}}_2$ need to satisfy   the average electrical power constraint, i.e.,
\begin{align}
{\eta _1}{\varepsilon _{\text{1}}}q_1^2 + {\eta _2}{\varepsilon _{\text{2}}}{\left| {{q_2}} \right|^2} \le {P_{\text{T}}},
\end{align}
 where  ${\eta_1}$ and ${\eta_2}$ denote the   efficiency of the   power amplifier of the  LiFi link and   WiFi link, respectively; ${\varepsilon _{\text{1}}}=\sum\limits_{k = 1}^M {{p_{1,k}}} x_{1,k}^2$, ${\varepsilon _{\text{2}}}=\sum\limits_{l = 1}^N {{p_{2,l}}}{{\left| {{x_{2,l}}} \right|}^2}$;
 and ${P_{\text{T}}}$ denotes the total  electrical power  threshold.
Moreover, for human eye safety  considerations, the power control over the LiFi signal also needs to meet the average optical
power   and   peak optical
power requirement: ${\mathbb{E}}\left\{ {{{q_1}}{x_1}} \right\} = q_1 \mu  \le {P_{\rm{o}}}$, and $q_1 A \le {P_{\rm{ins}}}$,
where  $\mu=\sum\limits_{k = 1}^M {{p_{1,k}}} {x_{1,k}}$; ${P_{\rm{o}}}$ and ${P_{\rm{ins}}}$ denote the maximum average   optical
power  threshold and the     instant optical
power threshold, respectively.

Let ${{\bf{g}}}\buildrel \Delta \over = \left[ {{{g}}_1}, {{{g}}_2} \right]^T$ denotes the channel vector, where  ${g}_1 $  and  ${g}_2 $  are the channel parameters of the LiFi link and   WiFi link, respectively.
{ Assume that the channel parameters $g_1$ and $g_2$ are quasi-static in the paper.}
Specifically,
${g}_1 $ is given by \cite{Ahmad}
\begin{align}
{g_1} = \frac{{\left( {m + 1} \right){A_1}}}{{2\pi d_1^2}}{\cos ^m}\left( \phi  \right)\cos \varphi {g_f}{g_c}\left(\varphi\right),
\end{align}
where $m$ is the order of the Lambertian emission;
$A_1$ is the detector area of PD receiver; $d_{1}$ is the distance between the LED and PD; ${\phi}$ and $\varphi$ are the radiance and incidence angle of the LiFi link; $g_f$ denotes the gain of the optical filter;  ${g_c}\left(\varphi\right)$  denotes the gain of optical concentrator, which is given as
\begin{align}
{g_c}\left(\varphi\right) = \left\{ {\begin{array}{*{20}{c}}
{ \frac{{{n^2}}}{{{{\sin }^2}\left( {{\Psi _c}} \right)}},}&{0 \le \varphi  \le {\Psi _c},}\\
{0,}&{{\rm{otherwise}},}
\end{array}} \right.
\end{align}
where $n$ is the refractive index, $\Psi _c$ represents the field-of-view (FoV) of the LiFi receiver,

As to the  WiFi link,   ${g}_2 $      is modeled by\cite{Wu}
\begin{align}
{g_{2}} = \left\{ {\begin{array}{*{20}{l}}
{{g_r}{{10}^{ - \left({{{L_{\rm{F}}}\left( {{d_{2}}} \right) + {L_\sigma }}}\right)/{{20}}}},{d_{2}} \le {d_{\rm{B}}}}\\
{{g_r}{{10}^{ - \left({{L_{\rm{F}}\left( {{d_{2}}} \right) + 35{{\log }_{10}}\left( {\frac{{{d_{2}}}}{{{d_{\rm{B}}}}}} \right) + {L_\sigma }}}\right)/{{20}}}},{d_{2}} > {d_{\rm{B}}},}
\end{array}} \right.
\end{align}
where $g_r=\left( { {\sqrt {\frac{K}{{K + 1}}} {e^{j\psi }} +  \sqrt{\frac{1}{{K + 1}}} {a}} } \right)$ denotes the small-scale fading gain,
 $a \sim {\cal C}{\cal N}\left( {0,1} \right)$; $\psi$ is the angle of arrival/departure of
the WiFi link;  $K$ denotes the Ricean $K$-factor;  $d_{2}$ denotes the distance between user and the RF antenna;  $ d_{\rm{B}}$ is  the breakpoint distance;
  ${L_\sigma }\sim {\cal N}\left( {0,\sigma^2} \right)$ denotes the shadow fading,   ${\sigma }=3$dB with ${d_2} \le {d_{\rm{B}}}$, ${\sigma }=5$dB with other conditions;
    ${L_{F}}\left( d_{2} \right)$ is the free space loss at the central carrier frequency $f_c$ as follows:
 \begin{align}
{L_{F}}\left( d_{2} \right)=20{\log _{10}}\left( d_{2} \right) + 20{\log _{10}}\left( {{f_c}} \right) - 147.5.
\end{align}

Let $y_1$ and $y_2$ denote the received signals    from  the  LiFi link and   WiFi link,  respectively,  which can be written in the vector form as
\begin{align}\label{receive}
\left[ \begin{array}{l}
 {y_1} \\
 {y_2} \\
 \end{array} \right] = \left[ {\begin{array}{*{20}{c}}
   {{g_1}{q_1}{x_1}}  \\
   {g_2^*{q_2}{x_2}}  \\
\end{array}} \right] + \left[ {\begin{array}{*{20}{c}}
   {{z_1}}  \\
   {{z_2}}  \\
\end{array}} \right],
 \end{align}
where      $z_1\sim {\cal N}\left( {0,\sigma^2 _1} \right)$ is the received real Gaussian  noise from  the LiFi link, and ${z_2}\sim {\cal C}{\cal N}\left( {0,\sigma ^2_2} \right)$  denotes the received complex Gaussian noise from  the   WiFi link. Then, the signal-to-noise ratio (SNR) of the LiFi  link and WiFi link can be defined as
\begin{align}
    \mathrm{SNR}_1 &\triangleq \frac{{g_1^2 {q_1^2}{\varepsilon _1}}}{{{B_1}\sigma _1^2}},~\mathrm{SNR}_2 \triangleq \frac{{{{\left| {{g_2}} \right|}^2} {\left|q_2\right|^2}{\varepsilon _2}}}{{{B_2}\sigma _2^2}}\label{snr2}.
\end{align}

Note that, for  the  aggregated LiFi-WiFi system with the finite-alphabet inputs,
the achievable rate    is  still unknown. To address this issue, we define the
      achievable rate   ${{{R}}_{{\rm{LiFi-WiFi}}}}$  as
\begin{align}\label{rate_LiFi_WiFi}
{R_{{\rm{LiFi - WiFi}}}} =  {{I}}\left( {{x_1},{x_2};{y_1},{y_2}} \right)= {R_{{\rm{LiFi}}}} + {R_{{\rm{WiFi}}}},
  \end{align}
   where
  ${{{R}}_{{\rm{LiFi}}}}\triangleq{{{I}}\left( {{x_1};{y_1}} \right)}
$  and ${{{R}}_{{\rm{WiFi}}}}\triangleq{{{I}}\left( {{x_2};{y_2}} \right)}$  denote    the achievable   rates   of  the   LiFi link and WiFi link, respectively.

{
    \emph{Proof:} 
    According to the mutual information, the upper bound of  $R_{\mathrm{LiFi-WiFi}}$ is given as
        \begin{subequations}
            \begin{align}
                R_{\mathrm{LiFi-WiFi}}&=I\left(x_1,x_2;y_1,y_2\right)\\
                &=h\left(y_1\right)+h\left(y_2|y_1\right)-h\left(y_1,y_2|x_1,x_2\right)\\
                &\leq I\left(x_1;y_1\right)+I\left(x_2;y_2\right)\label{eqn_total_rate_proof_c}\\
                &=R_{\mathrm{LiFi}}+R_{\mathrm{WiFi}},\label{upper_bound}
            \end{align}
        \end{subequations}
        where the equality   \eqref{eqn_total_rate_proof_c}  holds    if and only if $x_1$ and $x_2$ are independent. Thus, when $x_1$ and $x_2$ are independent, the upper bound \eqref{upper_bound} is achievable, i.e., the upper bound \eqref{upper_bound} is the channel capacity of the aggregated LiFi-WiFi system.
        In other words,  for the considered aggregated LiFi-WiFi system, the optimal inputs of  $x_1$ and $x_2$ are  independent, and   the achievable rate is ${R_{{\rm{LiFi - WiFi}}}} =  {{I}}\left( {{x_1},{x_2};{y_1},{y_2}} \right)= {R_{{\rm{LiFi}}}} + {R_{{\rm{WiFi}}}}$.  \qed
}

\textbf{Lemma 1:}  {\it
 With the finite-alphabet inputs and the given bandwidths $B_1$ and $B_2$, the achievable rates of the aggregated LiFi-WiFi system  ${{{R}}_{{\rm{LiFi}}}}$  and ${{{R}}_{{\rm{WiFi}}}}$ are respectively given as
  \begin{subequations}
\begin{align}
 &{{{R}}_{{\rm{LiFi}}}}=   - \frac{B_1}{{\ln 2}} - 2B_1\sum\limits_{k = 1}^{{M}} {{p_{1,k}}{\mathbb{E}_{{z_1}}}\left\{ {{{\log }_2}\sum\limits_{m = 1}^{{M}} {{p_{1,m}}\exp \left( {\Lambda_{k,m}} \right)} } \right\}},\label{rate_LiFi}\\
 &{{{R}}_{{\rm{WiFi}}}}=  - \frac{{B_{ 2}}}{{\ln 2}} - B_{ 2}\sum\limits_{l = 1}^{{N}} {{p_{2,l}}{\mathbb{E}_{{z_2}}}\left\{ {{{\log }_2}\sum\limits_{n = 1}^{{N}} {{p_{2,n}}\exp \left( \Gamma_{l,n} \right)} } \right\}},\label{rate_WiFi}
  \end{align}
\end{subequations}
where $\Lambda_{k,m}  \buildrel \Delta \over = { - \frac{{{{\left( {{g_1}{q_1}\left( {{x_{1,k}} - {x_{1,m}}} \right) + \sqrt {{B_1}}{z_1}} \right)}^2}}}{{2B_1\sigma _1^2}}} $,
$\Gamma_{l,n}  \buildrel \Delta \over = { - \frac{{{{\left| {{g_2^*}{q_2}\left( {{x_{2,l}} - {x_{2,n}}} \right) + \sqrt {{B_2}}{z_{2}}} \right|}^2}}}{{B_{ 2}\sigma _2^2}}}$.
\footnote{{
        The rate expression (10) is based on the unit bandwidth, while the Lemma 1 is for a given bandwidth. Thus, their units is different, i.e., bits/sec/Hz and bits/sec, respectively.
}}

}
\emph{Proof:} Please find the proof in Appendix \ref{app_lemma_1}.\qed

With Lemma 1, the achievable rate \eqref{rate_LiFi_WiFi} of the considered system can be obtained.

\section{Optimal   Discrete Constellation Inputs     of the Aggregated LiFi-WiFi System}

Based on the derived expression of  ${ R_{{\rm{LiFi - WiFi}}}}$, the following question is to optimize  the signal distributions and power allocation schemes for the two links to obtain the maximal achievable rate of the aggregated system.
  The optimization problem is formulated as
\begin{subequations}\label{max_rate_accuracy}
\begin{align}
 \mathop {\max }\limits_{q_1, q_2,\left\{ {{p_{1,k}}} \right\},\left\{ {{p_{2,l}}} \right\}} \quad &{ R_{{\rm{LiFi - WiFi}}}}\\
 {\rm{s}}{\rm{.t}}{\rm{.  }}  \quad\quad \quad  & {\eta_1} P_{\mathrm{e},1} q_1^2 + {\eta_2} P_{\mathrm{e},2} {\left| {{q_2}} \right|^2}\le {P_{\rm{T}}},\label{max_rate_accuracya}\\
 & q_1 \le \min \left( {{{{P_o}/\bar{\mu}}},{ {P_{\rm{ins}}}/A}{}} \right) ,\label{max_rate_accuracyb}\\
 &\sum\limits_{k = 1}^M {{p_{1,k}}} {x_{1,k}} \le \bar{\mu} , \sum\limits_{k = 1}^M {{p_{1,k}}} x_{1,k}^2 \le P_{{\rm{e}},1},\label{max_rate_accuracyd}\\
  &\sum\limits_{k = 1}^M {{p_{1,k}}}  = 1,{p_{1,k}} \ge 0,\forall k \in {\cal M}, \label{max_rate_accuracyc}\\
  &\sum\limits_{l = 1}^N {{p_{2,l}}{{\left| {{x_{2,l}}} \right|}^2}}  \le P_{{\rm{e}},2},\label{max_rate_accuracye} \\
     &\sum\limits_{l = 1}^N {{p_{2,l}}}  = 1,{p_{2,l}} \ge 0,\forall l \in {\cal N},\label{max_rate_accuracyf}
\end{align}
\end{subequations}
where \eqref{max_rate_accuracya} denotes the total average electrical power of two links; \eqref{max_rate_accuracyb} and \eqref{max_rate_accuracyd} denote the maximal optical power constraint, average optical power constraint, and average electrical power constraint on the LiFi link, respectively; \eqref{max_rate_accuracye} denotes the average electrical power constraint on the WiFi link; \eqref{max_rate_accuracyc} and \eqref{max_rate_accuracyf} denote the probability distribution constraint.


\begin{figure*}[!b]
    \hrulefill
    \normalsize
    \setcounter{mytempeqncnt}{\value{equation}}
    \setcounter{equation}{14}
\begin{align}\label{L_W}
    &{R_{{\rm{LiFi - WiFi}}}}=  - 2{B_1}\sum\limits_{k = 1}^M {{p_{1,k}}{\mathbb{E}_{{z_1}}}\left\{ {{{\log }_2}\sum\limits_{m = 1}^M {{p_{1,m}}\exp \left( { - \frac{{{{\left( {{g_1}\sqrt {{{\hat q}_1}} \left( {{x_{1,k}} - {x_{1,m}}} \right) + \sqrt {{B_1}} {z_1}} \right)}^2}}}{{2{B_1}\sigma _1^2}}} \right)} } \right\}} \nonumber\\
    &- {B_2}\sum\limits_{l = 1}^N {{p_{2,l}}{\mathbb{E}_{{z_2}}}\left\{ {{{\log }_2}\sum\limits_{n = 1}^N {{p_{2,n}}\exp \left( { - \frac{{{{\left| {\left| {{g_2}} \right|\sqrt {{{\hat q}_2}} \left( {{x_{2,l}} - {x_{2,n}}} \right) + \sqrt {{B_2}}{z_2}} \right|}^2}}}{{{B_2}\sigma _2^2}}} \right)} } \right\}}- \frac{{{B_1} + {B_2}}}{{\ln 2}}.
\end{align}
\end{figure*}
\setcounter{equation}{\value{mytempeqncnt}}

For problem \eqref{max_rate_accuracy}, the optimal phase of  $q_2$ is  the same  as that of $g_2$, which can be
proved by the counter-evidence method due to the property of the complex multiplication. Thus, the optimal $q_2$ can be written as
\begin{align}\label{q_2_0}
q_2=\sqrt {{{\hat q}_2}} \frac{{{g_2}}}{{\left| {{g_2}} \right|}},
\end{align}
where ${\hat q_2} \buildrel \Delta \over = {\left| {{q_2}} \right|^{\rm{2}}}$.
Furthermore, by substituting  \eqref{q_2_0} into \eqref{rate_LiFi_WiFi} and ${{{\hat q}_1}} \buildrel \Delta \over = q_1^2$,
   the  achievable rate  ${ R_{{\rm{LiFi - WiFi}}}}$ can be  rewrote as \eqref{L_W}.
\setcounter{equation}{15}

By defining ${\hat {\bf{x}}_1} \buildrel \Delta \over = {\left[ {{x_{1,1}}, \ldots ,{x_{1,{M}}}} \right]^{\rm{T}}}$ and ${{\bf{p}}_1} \buildrel \Delta \over = {\left[ {{p_{1,1}}, \ldots ,{p_{1,{M}}}} \right]^{\rm{T}}}$, constraints  \eqref{max_rate_accuracye} and  \eqref{max_rate_accuracyd} can be reformulated as
\begin{align}
&\Upsilon_1\triangleq\left\{\mathbf{p}\left|\mathbf{1}_M^{\mathrm{T}}\mathbf{p}=1, \mathbf{p}\succeq \mathbf{0},{\widehat{\mathbf{x}}_1^{\mathrm{T}}\mathbf{p}\leq\bar{\mu}},{\left(\widehat{\mathbf{x}}_1\odot\widehat{\mathbf{x}}_1\right)^{\mathrm{T}}\mathbf{p}\leq P_{e,1}}\right.\right\},\nonumber\\
&\mathbf{p}_1\in\Upsilon_1.
\end{align}
Similarly, by defining ${\hat {\bf{x}}_2} \buildrel \Delta \over = {\left[ {{x_{2,1}} \ldots {x_{2,{N}}}} \right]^{\rm{T}}}$ and ${{\bf{p}}_2} \buildrel \Delta \over = {\left[ {{p_{2,1}} \ldots {p_{2,{N}}}} \right]^{\rm{T}}}$, constraints
\eqref{max_rate_accuracyc} and \eqref{max_rate_accuracyf} can be reformulated as
\begin{align}
\Upsilon_2&\triangleq\left\{\mathbf{p}\left|\mathbf{1}_N^{\mathrm{T}}\mathbf{p}=1, \mathbf{p}\succeq \mathbf{0},{\left(\widehat{\mathbf{x}}_2\odot\widehat{\mathbf{x}}_2\right)^{\mathrm{T}}\mathbf{p}\leq P_{e,2}}\right.\right\},\mathbf{p}_2\in\Upsilon_2.
\end{align}

Next, we  introduce   auxiliary variables, i.e.,
\begin{subequations}
\begin{align}
 &{\bf{w}} \buildrel \Delta \over = {\left[ { {{{\log }_2}{\bf{p}}_1^{\rm{T}}{{{\bf{\hat w}}}_1}} , \ldots , {{{\log }_2}{\bf{p}}_1^{\rm{T}}{{{\bf{\hat w}}}_M}}}  \right]^{\rm{T}}},\\
 &{{{\bf{\hat w}}}_k} \buildrel \Delta \over = {\left[ {{{\hat w}_{k,1}}, \ldots ,{{\hat w}_{k,M}}} \right]^{\rm{T}}},\\
 &{{\hat w}_{k,m}} \buildrel \Delta \over = \exp \left( { - \frac{{{{\left( {{g_1}\sqrt {{{\hat q}_1}} \left( {{x_{1,k}} - {x_{1,m}}} \right) + \sqrt {{B_1}}{z_1}} \right)}^2}}}{{2{B_1}\sigma _1^2}}} \right),\\
 &{\bf{r}} \buildrel \Delta \over = {\left[ { {{{\log }_2}{\bf{p}}_2^{\rm{T}}{{{\bf{\hat r}}}_1}} , \ldots , {{{\log }_2}{\bf{p}}_2^{\rm{T}}{{{\bf{\hat r}}}_N}} } \right]^{\rm{T}}},\\
 &{{{\bf{\hat r}}}_l} \buildrel \Delta \over = {\left[ {{{\hat r}_{l,1}}, \ldots ,{{\hat r}_{l,N}}} \right]^{\rm{T}}}, \\
 &{{\hat r}_{l,n}} \buildrel \Delta \over = \exp \left( { - \frac{{{{\left| {\left| {{g_2}} \right|\sqrt {{{\hat q}_2}} \left( {{x_{2,l}} - {x_{2,n}}} \right) + \sqrt {{B_2}}{z_2}} \right|}^2}}}{{{B_2}\sigma _2^2}}} \right),
\end{align}
\end{subequations}
where $k$ and $m \in {\cal M}$,  $l$ and $n  \in {\cal N}$,
Then, we rewrite  the achievable rate  ${ R_{{\rm{LiFi - WiFi}}}}$   as
\begin{align}\label{rate_equ}
 {R_{{\rm{LiFi - WiFi}}}}= &- \frac{{{B_1} + {B_2}}}{{\ln 2}}
 - 2{B_1}{\mathbb{E}_{{z_1}}}\left\{{\bf{p}}_1^{\rm{T}}{\bf{w}}\right\}- {B_2}{\mathbb{E}_{{z_2}}}\left\{{\bf{p}}_2^{\rm{T}}{\bf{r}}\right\}.
  \end{align}

Based on the above definitions, problem \eqref{max_rate_accuracy}  can be equivalent to a compact from as
\begin{subequations}\label{max_rate_accuracy_2}
\begin{align}
 \mathop {\min }\limits_{\hat q_1, \hat q_2,{\bf{p}}_1,{\bf{p}}_2} \quad &  2{B_1}{\mathbb{E}_{{z_1}}}\left\{{\bf{p}}_1^{\rm{T}}{\bf{w}}\right\} + {B_2}{\mathbb{E}_{{z_2}}}\left\{{\bf{p}}_2^{\rm{T}}{\bf{r}}\right\} \label{max_rate_accuracy_2a}\\
 {\rm{s}}{\rm{.t}}{\rm{.  }} \quad\quad &{\eta_1} P_{\mathrm{e}, 1}\hat q_1 + {\eta_2} P_{\mathrm{e}, 2}\hat q_2 \le {P_{\rm{T}}},\label{max_rate_accuracy_2b}\\
 & \hat q_1 \le \tau^2,\label{max_rate_accuracy_2c}\\
 &{{\bf{p}}_1 \in {\Upsilon _1}},{{\bf{p}}_2 \in {\Upsilon _2}},\label{max_rate_accuracy_2d}
\end{align}
\end{subequations}
where $\tau \buildrel \Delta \over =\min \left( {{{{P_o}/\bar{\mu}}},{ {P_{\rm{ins}}}/A}{}} \right)$.
Note that in problem \eqref{max_rate_accuracy_2},
  the power allocation variables $ {{\hat { q}_1}}$ and $ {{\hat { q}_2}}$  are only contained in
constraints \eqref{max_rate_accuracy_2b} and \eqref{max_rate_accuracy_2c}, while
the distribution  variables  ${\bf{p}}_1$ and ${\bf{p}}_2$ are only contained in the constraint \eqref{max_rate_accuracy_2d}.
Therefore, the optimization problem (18) can be decomposed into two sub-problems solved  alternately until the objective function converges
: \emph{power allocation sub-problem 1}: optimizing $ {{\hat { q}_1}}$ and $ {{\hat { q}_2}}$ with given ${\bf{p}}_1$ and ${\bf{p}}_2$, and  \emph{probability distribution sub-problem 2}: optimizing ${\bf{p}}_1$ and ${\bf{p}}_2$ with given $ {{\hat { q}_1}}$ and $ {{\hat { q}_2}}$.
Next, we will present the solutions to these two sub-problems.

\subsection{\textbf{Power Allocation Sub-Problem} }

With given  ${\bf{p}}_1$, ${\bf{p}}_2$,  problem \eqref{max_rate_accuracy_2} can be simplified to an optimal power allocation problem as
\begin{subequations}\label{max_rate_accuracy_power}
\begin{align}
 \mathop {\min }\limits_{\hat q_1, \hat q_2} \quad  & h\left( {\hat q_1},  {\hat q_2}  \right) \\
 {\rm{s}}{\rm{.t}}{\rm{.  }} \quad &{\eta_1} P_{\mathrm{e},1}\hat q_1  + {\eta_2} P_{\mathrm{e},2}{\hat q_2}\le {P_{\rm{T}}}, \label{max_rate_agg_b}\\
 & {\hat q_1} \le \tau^2,\label{max_rate_agg_c}
\end{align}
\end{subequations}
where $h\left( {\hat q_1},  {\hat q_2}  \right) \buildrel \Delta \over = 2{B_1}{\mathbb{E}_{{z_1}}}\left\{{\bf{p}}_1^{\rm{T}}{\bf{w}}\right\} + {B_2}{\mathbb{E}_{{z_2}}}\left\{{\bf{p}}_2^{\rm{T}}{\bf{r}}\right\}$.

  Note that {$-\frac{\left(g_1 \sqrt{\hat{q}_1}\left(x_{1,k}-x_{1,m}\right)+\sqrt{B_1} z_1\right)^2}{2 B_1 \sigma_1^2}$} and {$-\frac{\left|\left|g_2\right|\sqrt{\hat{q}_2}\left(x_{2,\ell}-x_{2,n}\right)+\sqrt{B_2} z_2\right|^2}{B_2 \sigma_2^2}$} are  convex  over $\hat q_1$ and $\hat q_2$, respectively.
Furthermore, since $\log \sum\limits_i {{e^{{f_i}\left( x \right)}}} $
 is convex as long as ${{f_i}\left( x \right)}$
 is convex\cite{Boyd},      problem \eqref{max_rate_accuracy_power} is convex over $\hat q_1$ and $\hat q_2$.
To obtain  the  optimal power allocation $\hat q_1$ and $\hat q_2$, we write the Lagrangian function  of problem \eqref{max_rate_accuracy_power} as
\begin{align}
{\cal L}\left({\hat q_1},  {\hat q_2}, \gamma,\nu  \right) =& h\left( {\hat q_1},  {\hat q_2}  \right) + \gamma \left( {{\eta _1}{{P_{e,1}}}\hat q_1 + {\eta _2}{{P_{e,2}}}{\hat q_2} - {P_{\rm{T}}}} \right)\nonumber\\
&~ + \nu \left( {{\hat q_1} - \tau^2} \right),
\end{align}
where $ \gamma \ge 0 $ and $ \nu \ge 0 $ are the Lagrangian multipliers associated with constraints
\eqref{max_rate_agg_b} and \eqref{max_rate_agg_c}, respectively. Then, the Karush-Kuhn-Tucker (KKT) conditions of problem \eqref{max_rate_accuracy_power} are
\begin{subequations}\label{KKT}
\begin{align}
 &\frac{{\partial h\left( {\hat q_1},  {\hat q_2}  \right)}}{{\partial {\hat q_1}}} + \gamma {\eta _1}P_{\mathrm{e},1} + \nu  = 0,\\
 &\frac{{\partial h\left( {\hat q_1},  {\hat q_2}  \right)}}{{\partial {\hat q_2}}} +  \gamma {\eta _2}P_{\mathrm{e},2} = 0,\\
 &\gamma \left( {{\eta _1}P_{\mathrm{e},1}\hat q_1 + {\eta _2}P_{\mathrm{e},2}{\hat q_2} - {P_{\rm{T}}}} \right) = 0,~\gamma  \ge 0,\\
 &\nu \left( {{\hat q_1} - \tau^2} \right) = 0, ~\nu  \ge 0.
\end{align}
\end{subequations}

   Furthermore, based on the definition of ${h\left( {{{\hat q}_1},{{\hat q}_2}} \right)}$, we have
\begin{align}
   \frac{{\partial h\left( {{{\hat q}_1},{{\hat q}_2}} \right)}}{{\partial {{\hat q}_1}}} &=  - \frac{{g_1^2{\varepsilon _1}}}{{{B_1}\sigma _1^2}}\frac{{\partial {I}\left( {{x_1};{y_1}} \right)}}{{\partial {\rm{SN}}{{\rm{R}}_{\rm{1}}}}},\nonumber\\
   \frac{{\partial h\left( {{{\hat q}_1},{{\hat q}_2}} \right)}}{{\partial {{\hat q}_2}}} &=  - \frac{{{{\left| {{g_2}} \right|}^2}{\varepsilon _2}}}{{{B_2}\sigma _2^2}}\frac{{\partial {\rm{I}}\left( {{x_2};{y_2}} \right)}}{{\partial {\rm{SN}}{{\rm{R}}_2}}}.
 \end{align}
Moreover, according to relationship between the mutual information and the MMSE \cite{Shamai}, we have
\begin{align}\label{MMSE}
    \begin{aligned}
        \frac{{\partial {{I}}\left( {{x_1};{y_1}} \right)}}{{\partial {\rm{SNR}_1}}} &= \frac{1}{2}{\rm{MMS}}{{\rm{E}}_1}\left( {{\rm{SNR}}_1} \right),\\
        \frac{{\partial {{I}}\left( {{x_2};{y_2}} \right)}}{{\partial {\rm{SN}}{{\rm{R}}_2}}} &= {\rm{MMSE}}_2\left( {{\rm{SN}}{{\rm{R}}_2}} \right),
    \end{aligned}
\end{align}
where $\mathrm{MMSE}_i\left(\mathrm{SNR}_i\right)\triangleq\mathbb{E}\left\{\left|x_i-\hat{x}_i\right|\right\}$ denotes the MMSE between $x_i$ and $\hat{x}_i$, $i = 1,2$, and $\hat{x}_i$ is the conditional mean of the MMSE estimate of $x_i$, i.e. $\hat{x}_1\triangleq\mathbb{E}\left\{x_1|y_1=g_1 q_1 x_1 + z_1\right\}$ and $\hat{x}_2\triangleq\mathbb{E}\left\{x_2|y_2=g_2^* q_2 x_2 + z_2\right\}$.
Note that   the calculation of the above MMSE involves non-holonomic function{, i.e., it is non-trivial to compute the exact MMSE for an arbitrary input distribution. Meanwhile, it is an opening problem to find the lower and upper bounds of the MMSE for the discrete inputs \cite{Cao}. Thus, most} of existing works adopt numerical methods such as Monte Carlo integral  to obtain the results, which may lead  to sub-optimality and high complexity. To overcome   this issue,
we approximate  the  MMSEs    by a scaled
linear MMSE upper bound \cite{Cao}, i.e.,
 \begin{align}
     \begin{aligned}
         \rm{MMSE}_1 \left(\rm{SNR_1}\right)  &\approx  \rm{LMMSE}_{\varepsilon _1}\left(\rm{SNR_1}\right),\\
         \rm{MMSE}_2 \left(\rm{SNR_2}\right)  &\approx   \rm{LMMSE}_{\varepsilon _2}\left(\rm{SNR_2}\right),
     \end{aligned}
\end{align}
where ${\rm{LMMSE}}_t\left(x \right)=\frac{t}{1+tx}$. {Although the approximation based on  LMMSE, like a continuous Gaussian MMSE, would lead to the optimality loss, it can provide a fast acceptable closed-form method.}
Then,   \eqref{MMSE} can be reformulated as
\begin{align}\label{MMSE1}
\begin{aligned}
    \frac{{\partial {{I}}\left( {{x_1};{y_1}} \right)}}{{\partial {\rm{SNR}_1}}} &= \frac{1}{2}{\rm{LMMSE}}_{\varepsilon _1}\left(\rm{SNR_1}\right),\\
    \frac{{\partial {{I}}\left( {{x_2};{y_2}} \right)}}{{\partial {\rm{SN}}{{\rm{R}}_2}}} &= { \rm{LMMSE}}_{\varepsilon _2}\left(\rm{SNR_2}\right).
\end{aligned}
\end{align}
With  \eqref{snr2} and \eqref{MMSE1}, we  arrive at
\begin{align}\label{MMSE_q1}
\begin{aligned}
    \frac{{\partial h\left( {\hat q_1},  {\hat q_2}  \right)}}{{\partial {\hat q_1}}} &= -\frac{{g_1^2{\varepsilon _1}}}{{2{B_1}\sigma _1^2}}{\rm{LMMSE}}_{\varepsilon _1}\left( {\frac{{g_1^2 \hat q_1{\varepsilon _1}}}{{{B_1}\sigma _1^2}}} \right),\\
    \frac{{\partial h\left( {\hat q_1},  {\hat q_2}  \right)}}{{\partial {\hat q_2}}} &= -\frac{{{{\left| {{g_2}} \right|}^2}{\varepsilon _2}}}{{{B_2}\sigma _2^2}} {\rm{LMMSE}}_{\varepsilon _2}\left( {\frac{{{{\left| {{g_2}} \right|}^2}{\hat q_2}{\varepsilon _2}}}{{{B_2}\sigma _2^2}}} \right).
\end{aligned}
\end{align}

Substituting equation \eqref{MMSE_q1} into equation \eqref{KKT}, we have
\begin{subequations}\label{KKT_q1}
\begin{align}
&\frac{{g_1^2{\varepsilon _1}}}{{2{B_1}\sigma _1^2}}{\rm{LMMSE}}_{\varepsilon _1}\left( {\frac{{g_1^2 \hat q_1{\varepsilon _1}}}{{{B_1}\sigma _1^2}}} \right) - \gamma {\eta _1}P_{\mathrm{e},1} - \nu  = 0,\\
&\frac{{{{\left| {{g_2}} \right|}^2}{\varepsilon _2}}}{{{B_2}\sigma _2^2}}{\rm{LMMSE}}_{\varepsilon _2}\left( {\frac{{{{\left| {{g_2}} \right|}^2}{\hat q_2}{\varepsilon _2}}}{{{B_2}\sigma _2^2}}} \right) -  \gamma {\eta _2}P_{\mathrm{e},2} = 0.
\end{align}
\end{subequations}

Then, $\widehat{q}_1$ and $\widehat{q}_2$ can be derived through solving the equation \eqref{KKT_q1} as
\begin{subequations}
\begin{align}
&{{\hat q}_1}=\frac{{{B_1}\sigma _1^2}}{{g_1^2{\varepsilon _1}}}\rm{LMMSE}_{\varepsilon _1}^{{\rm{ - 1}}}\left( {\frac{{2{B_1}\sigma _1^2\left( {\gamma {\eta _1}P_{\mathrm{e},1}{\rm{ + }}\nu } \right)}}{{g_1^2{\varepsilon _1}}}} \right),\label{solve_q1_a}\\
&{{\hat q}_2}=\frac{{{B_2}\sigma _2^2}}{{{{\left| {{g_2}} \right|}^2}{\varepsilon _2}}}\rm{LMMSE}_{\varepsilon _2}^{{\rm{ - 1}}}\left( {\frac{{{B_2}\sigma _2^2\gamma {\eta _2}P_{\mathrm{e},2}}}{{{{\left| {{g_2}} \right|}^2}{\varepsilon _2}}}} \right),\label{solve_q1_b}
\end{align}
\end{subequations}
where ${\rm{LMMSE}}_t^{ - 1}\left( x \right) = \frac{1}{x} - \frac{1}{t}$.
The Lagrangian multipliers $ \gamma $ and $  \nu  $ can be solved by the  Water-filling (WF)  Method. The detailed algorithm is given in Algorithm 1.

\begin{algorithm}[htbp]
  \caption{  Water-filling (WF)  Method for Power Allocation Sub-Problem \eqref{max_rate_accuracy_power}.}
  \label{alg:Framwork}
  \begin{algorithmic}[1]
    \State  Given  $\zeta \ge 0$, $\gamma  \in \left[ 0,\hat \gamma \right]$ ;
    \State \textbf{Initialization}: ${\gamma _{\min }}=0$, ${\gamma _{\max }}=\hat \gamma $;
      \State \textbf{repeat}.
         \State \quad Set $\gamma  \leftarrow \left( {{\gamma _{\min }} + {\gamma _{\max }}} \right)/2$;
     \State   \quad Obtain     $\hat q_2$ by \eqref{solve_q1_b};
     \State \quad Find the minimum $\nu $ satisfying the constraint  ${\hat q_1} \le \tau^2 $;
      \State \quad Obtain    $\hat q_1$ by \eqref{solve_q1_a};
       \State \quad If $ {\eta _1}P_{\mathrm{e},1} \hat q_1 + {\eta _2}P_{\mathrm{e},2}{\hat q_2} \le {P_{\rm{T}}}  $,
       set ${\gamma _{\max }} \leftarrow \gamma  $;    otherwise, ${\gamma _{\min }} \leftarrow \gamma  $;
    \State \textbf{until} $\left| {{\nu _{\max }} - {\nu _{\min }}} \right| \le \zeta $.
        \State
    \Return $\hat q_1$ and  $\hat q_2$;
  \end{algorithmic}
\end{algorithm}

\subsection{\textbf{Probability Distribution Sub-Problem}}

When $\hat q_1$ and $\hat q_2$ are given,   problem \eqref{max_rate_accuracy_2} can be reformulated as
\begin{subequations}\label{max_rate_accuracy_pro}
\begin{align}
\mathop {\min }\limits_{{\bf{p}}_1,{\bf{p}}_2} \quad  &  2{B_1}{\mathbb{E}_{{z_1}}}\left\{{\bf{p}}_1^{\rm{T}}{\bf{w}}\right\} + {B_2}{\mathbb{E}_{{z_2}}}\left\{{\bf{p}}_2^{\rm{T}}{\bf{r}}\right\}\label{max_rate_accuracy_pro_obj}\\
 {\rm{s}}{\rm{.t}}{\rm{.  }}\quad  &{{\bf{p}}_1 \in {\Upsilon _1}},~{{\bf{p}}_2 \in {\Upsilon _2}},
\end{align}
\end{subequations}
which is a convex optimization problem with     two  variables  ${\bf{p}}_1$ and ${\bf{p}}_2$.
However, there is no
 analytical expression of the   objective function, which prevents us from
calculating the optimal  probability distributions.

To overcome this bottleneck, we adopt the inexact   gradient
descent method \cite{Ma_ISJ_2020} to calculate  the optimal  probability distribution.
{ Besides, the Blahut-Arimoto algorithm \cite{Favano2021,Dytso2021,Dauwels2006} can also solve this problem.}
 Let $\phi_1 \left( {{{\bf{p}}_1}} \right) \buildrel \Delta \over=  2{B_1}{\mathbb{E}_{{z_1}}}\left\{{\bf{p}}_1^{\rm{T}}{\bf{w}}\right\}$ and ${\nabla }\phi_1\left( {{{{\bf{ p}}}_{1}}} \right)$
denote the its gradient as
 \begin{align}\label{rate_LiFi_gradient}
{\nabla }\phi_1\left( {{{{\bf{ p}}}_{1}}} \right) &=2{B_1}{\mathbb{E}_{{z_1}}}\left\{{{\bf{w}} + {\bf{W}}{{\bf{p}}_1}} \right\}\nonumber\\
&=2{B_1} \int_{ - \infty }^\infty  {{f_{{z_1}}}\left( {{z_1}} \right)\left( {{\bf{w}} + {\bf{W}}{{\bf{p}}_1}} \right)} dz_1,
\end{align}
where ${ {\bf{W}} } \buildrel \Delta \over ={\left[ { {{W}}_{i,j} } \right]}$, $ { {{W}}_{i,j} } \buildrel \Delta \over = \frac{{{\bf{\hat w}}_j^{\rm{T}}{{\bf{e}}_i}}}{{{\bf{\hat w}}_j^{\rm{T}}{{\bf{p}}_1}\ln 2}}$,  ${{\bf{e}}_i}$ denotes the unit vector in which the $i$th element is 1 and the other elements are zeros, and ${f_{{z_1}}}\left( {{z_1}} \right) \buildrel \Delta \over = \frac{1}{{\sqrt {2\pi } {\sigma _1}}}\exp \left( { - \frac{{z_1^2}}{{\sigma _1^2}}} \right)$ denotes the probability density function of $z_2$.

 Since neither
  ${\phi _j}\left( {{{{\bf{p}}}_j}} \right)$ nor $\nabla {\phi _j}\left( {{{\bf{\hat{p}}}_j}} \right)$ has a closed-form expression,   we perform the truncation of the integration from infinity to an finite interval.
 More  specifically, let $\left[ { - {\tau _1},{\tau _1}} \right]$ and $\left[ { - {\tau _2},{\tau _2}} \right]$ denote the integration intervals of $\phi_1\left( {{{{\bf{ p}}}_{1}}} \right)$ and ${\nabla } \phi_1\left( {{{{\bf{ p}}}_{1}}} \right)$. Then, let  $\hat \phi_1\left( {{{{\bf{ p}}}_{1}}} \right)$ and ${\nabla } \hat \phi_1\left( {{{{\bf{ p}}}_{1}}} \right)$ denote the approximation of $\phi_1\left( {{{{\bf{ p}}}_{1}}} \right)$ and ${\nabla } \phi_1\left( {{{{\bf{ p}}}_{1}}} \right)$ respectively, which are given by
\begin{subequations}\label{gra_app_1}
\begin{align}
&\hat \phi_1 \left( {{{\bf{p}}_1}} \right) \buildrel \Delta \over=  2{B_1} \int_{ - {\tau _1} }^{\tau _1}{f_{{z_1}}}\left( {{z_1}} \right){\bf{p}}_1^{\rm{T}}{\bf{w}}dz,\\
&{\nabla }\hat \phi_1\left( {{{{\bf{ p}}}_{1}}} \right) \buildrel \Delta \over= 2{B_1}\int_{ - {\tau _2} }^{\tau _2}  {{f_{{z_1}}}\left( {{z_1}} \right)\left( {{\bf{w}} + {\bf{W}}{{\bf{p}}_1}} \right)} dz,
\end{align}
\end{subequations}
where ${\tau_1} > 0$ and ${\tau_2} > 0$.

  With the approximated objective function and its gradient, i.e., $\hat \phi_1\left( {{{{\bf{ p}}}_{1}}} \right)$ and ${\nabla } \hat \phi_1\left( {{{{\bf{ p}}}_{1}}} \right)$,  we adopt the    gradient projections method to solve problem \eqref{max_rate_accuracy_pro}. Specifically, let  $ {{{\bf{p}}^{\left[ {i} \right]}_{1}}}$ denote the $n$th iteration  feasible point.
With  inexact gradient ${\nabla } \hat \phi_1\left( {{{{\bf{ p}}}_{1}}} \right)$, the gradient projection iteration between $ {{{\bf{p}}^{\left[ {i} \right]}_{1}}}$ and  $ {{{\bf{p}}^{\left[ {i+1} \right]}_{1}}}$ is given by
\begin{align}
{{{\bf{p}}^{\left[ {i+1} \right]}_{1}}}  = {\rm{Pro}}{{\rm{j}}_{ {\Upsilon _1}}}\left( {{\bf{p}}_1^{\left[ i \right]} - \alpha _1^{\left[ i \right]}{\nabla }\hat \phi_1\left( {\bf{p}}_1^{\left[ i \right]}  \right)} \right),
\end{align}
where $\alpha _1^{\left[ i \right]}$ denotes the $i$th-iteration step size, ${\rm{Pro}}{{\rm{j}}_{ {\Upsilon _1}}} \left( {\bf{\hat x}}\right)$ denotes the   projection of ${\bf{ x}}$ onto ${\Upsilon _1}$:
\begin{align}\label{proj_d}
{{\rm{Pro}}{{\rm{j}}_{{\Upsilon _1}}}\left( {{{ {\bf{{x}}}}}} \right) = \left\{ {\begin{array}{*{20}{l}}
{{{ {\bf{{x}}}}},{\rm{if}}\;{{ {\bf{{x}}}}} \in {\Upsilon _1},}\\
{\mathop {\arg \min }\limits_{{{{{\bf{{\hat{x}}}}}}} \in {\Upsilon _1}} \left\| {{{ {\bf{{x}}}}} - {{{{\bf{{\hat{x}}}}}}}} \right\|^2,{\rm{otherwise}}.}
\end{array}} \right.}
\end{align}

The details of the  inexact gradient descent method are listed in Algorithm 2, and  the distribution of LiFi  link ${\bf{p}}_1$ can be obtained.
The optimal distribution of  WiFi  link ${\bf{p}}_2$ can be   obtained analogously. Thus, we
omit the detailed derivations for brevity. { The optimal probability distribution $\mathbf{p}_1$, \(\mathbf{p}_2\) can be achieved by the distribution matching (DM) \cite{Schulte2016,Chen2017,Buchali2016}.}

In summary,  the  achievable rate maximization problem \eqref{max_rate_accuracy}
can be solved by Algorithm 3    , therefore the maximal achievable rate $ R_{{\rm{LiFi-WiFi}}}$ can be achieved.

\begin{algorithm}[htbp]
    \caption{ Inexact Gradient Descent Method for Probability Distribution Sub-Problem\eqref{max_rate_accuracy_pro}}
    \label{alg:Framwork_O_1}
    \begin{algorithmic}[1]
        \State \textbf{Initialization}: Given  $\delta  \ge 0$, set $i=1$, and choose ${{{{\bf{p}}}^{\left[ {i} \right]}_{1}}} \in {\Upsilon _1}$.
        \State  \textbf{repeat}
        \State \quad  $i \leftarrow i+1$.
        \State  \quad Update $\hat \phi_1\left( {{{{\bf{p}}}^{\left[ {i-1} \right]}_{1}}} \right)$ and ${\nabla } \hat \phi_1\left( {{{{\bf{p}}}^{\left[ {i-1} \right]}_{1}}} \right)$ by \eqref{gra_app_1}.
        \State   \quad Compute the step size  $\alpha _1^{\left[ i1- \right]}$  by Armijo rule\cite{Boyd}.
        \State \quad Update  ${{{\bf{p}}^{\left[ {i} \right]}_{1}}}  = {\rm{Pro}}{{\rm{j}}_{ {\Upsilon _1}}}\left( {{\bf{p}}_1^{\left[ i-1 \right]} - \alpha _1^{\left[ i -1\right]}{\nabla }\hat \phi_1\left( {\bf{p}}_1^{\left[ i-1 \right]}  \right)} \right)$.
        \State \textbf{until}  $\left\| {{{\bf{p}}^{\left[ {i} \right]}_{1}}}-{{{\bf{p}}^{\left[ {i-1} \right]}_{1}}} \right\| \le\delta  $.
        \State
        \Return ${{{\bf{p}}_{1}}} ={{{\bf{p}}^{\left[ {i} \right]}_{1}}} $.
    \end{algorithmic}
\end{algorithm}

\begin{algorithm}[htbp]
  \caption{  Optimal   discrete constellation inputs  of problem \eqref{max_rate_accuracy_2}}
  \label{alg:Framwork_O_2}
  \begin{algorithmic}[1]
    \State \textbf{Initialization}: Given  $\xi  \ge 0$, set $k=1$, and choose ${{{{\bf{p}}}_{1}^{\left[ 1 \right]}}} \in {\Upsilon _1}$, ${{{{\bf{p}}}_{2}^{\left[ 1 \right]}}} \in {\Upsilon _2}$.
    \State  Update ${{\hat q}_{1}^{\left[ 1 \right]}}$ and ${{\hat q}_{2}^{\left[ 1 \right]}}$ by  Algorithm 1.
    \State Obtain $R_{\mathrm{LiFi-WiFi}}^{\left[1\right]}$ by substituting $\hat{q}_1^{\left[1\right]}$, $\hat{q}_2^{\left[1\right]}$, $\mathbf{p}_1^{\left[1\right]}$ and $\mathbf{p}_2^{\left[1\right]}$ into formulation \eqref{rate_equ}.
      \State  \textbf{repeat}
      \State \quad $k \leftarrow k+1$.
        \State \quad Update ${{\bf{ p}}_{1}^{\left[ k \right]}}$ and  ${{\bf{ p}}_{2}^{\left[ k  \right]}}$ based on  Algorithm 2  with given ${\hat q}_{1}^{\left[ k-1  \right]}$ and ${\hat q}_{2}^{\left[ k -1 \right]}$.
     \State \quad Update ${\hat q}_{1}^{\left[ k  \right]}$ and ${\hat q}_{2}^{\left[ k  \right]}$ by  Algorithm 1 with given ${{\bf{ p}}_{1}^{\left[ k \right]}}$ and  ${{\bf{ p}}_{2}^{\left[ k  \right]}}$.
        \State \quad Obtain $ R^{\left[ k \right]}_{{\rm{LiFi-WiFi}}}$ by substituting ${\hat q}^{\left[ k \right]}_{1}$ , ${\hat q}^{\left[ k \right]}_{2}$, ${{\bf{ p}}^{\left[ k \right]}_{1}}$ and  ${{\bf{ p}}^{\left[ k \right]}_{2}}$ into formulation \eqref{rate_equ}.
     \State  \textbf{until} $\left| { R^{\left[ k \right]}_{{\rm{LiFi-WiFi}}} - R^{\left[ k -1\right]}_{{\rm{LiFi-WiFi}}}} \right| \le \xi $.
    \State
    \Return ${{{\bf{p}}_{1}}} ={{{\bf{p}}^{\left[ k \right]}_{1}}} $, ${{{\bf{p}}_{2}}} ={{{\bf{p}}^{\left[ k \right]}_{2}}} $, ${\hat q}_1= {\hat q}^{\left[ k \right]}_1$ and ${\hat q}_2= {\hat q}^{\left[ k \right]}_2$,$ R_{{\rm{LiFi-WiFi}}}=R^{\left[ k-1 \right]}_{{\rm{LiFi-WiFi}}}$.
  \end{algorithmic}
\end{algorithm}

\section{Optimal Discrete Constellation Input Distributions based on    Lower Bounds and Upper Bounds}

Recall that, neither  the achievable rate \eqref{rate_LiFi} nor \eqref{rate_WiFi} is in a closed-form
expression, and thus the  calculation of \eqref{rate_LiFi_WiFi}  is computationally inefficient.
 To reduce the computational
complexity,
we   may replace the objective function with an explicit expression. Thus, we turn to the capacity bound of  LiFi link and WiFi link.

\textbf{Lemma 2:}  {\it With the  finite-alphabet inputs,
the closed-form upper bound ${R^{\rm{U}}_{{\rm{LiFi}}}}$ and lower bound ${R^{\rm{L}}_{{\rm{LiFi}}}}$ of the LiFi link achievable rate  are respectively given as
 \begin{subequations}\label{rate_up_low_L}
\begin{align}
&{R^{\rm{U}}_{{\rm{LiFi}}}} =
 -2B_{1} \sum\limits_{k = 1}^{{M}} {{p_{1,k}}{{\log }_2}\sum\limits_{m = 1}^{{M}} {{p_{1,m}}} \exp \left( 2\hat \Lambda_{k,m}  \right)},\label{rate_LiFi_up}\\
 &{R^{\rm{L}}_{{\rm{LiFi}}}}  =
B_{1}- \frac{B_{1}}{{\ln 2}}    - 2B_{1}\sum\limits_{k = 1}^{{M}} {{p_{1,k}}{{\log }_2}\sum\limits_{m = 1}^{{M}} {{p_{1,m}}} \exp \left( \hat \Lambda_{k,m}  \right)},\label{rate_LiFi_low}
  \end{align}
  \end{subequations}
  where $\hat \Lambda_{k,m}  \buildrel \Delta \over = { - \frac{{{{\left( {{x_{1,k}} - {x_{1,m}}} \right)}^2}{{\left( {{g_1}{q_1}} \right)}^2}}}{{4B_{1}\sigma _1^2}}}$.
  }

  \emph{Proof:} Please find the proof in  Appendix \ref{app_lemma_2}.\qed

Furthermore, we   develop  the closed-form upper bound and lower bound of the achievable rate for WiFi link  ${{{R}}_{{\rm{WiFi}}}}$.

\textbf{Lemma 3:}  {\it
The upper bound ${R^{\rm{U}}_{{\rm{WiFi}}}}$ and lower bound ${R^{\rm{L}}_{{\rm{WiFi}}}}$ of the achievable rate for the   WiFi link  are respectively given as
\begin{subequations}
\begin{align}
 &{R^{\rm{U}}_{{\rm{WiFi}}}} = -  B_{2}\sum\limits_{l = 1}^{{N}} {{p_{2,l}}{{\log }_2}\sum\limits_{n = 1}^{{N}} {{p_{2,n}}} \exp \left( 2\hat\Gamma _{l,n} \right)},\label{rate_WiFi_up}\\
 &{R^{\rm{L}}_{{\rm{WiFi}}}} =B_{2}- \frac{B_{2}}{{\ln 2}}     - B_{2} \sum\limits_{l = 1}^{{N}} {{p_{2,l}}{{\log }_2}\sum\limits_{n = 1}^{{N}} {{p_{2,n}}} \exp \left( \hat\Gamma _{l,n} \right)}\label{rate_WiFi_L1_1},
  \end{align}
  \end{subequations}
  where $\hat\Gamma _{l,n}  \buildrel \Delta \over = { - \frac{{{{\left| {{x_{2,l}} - {x_{2,n}}} \right|}^2}{{\left| {g_2^*{q_2}} \right|}^2}}}{{2B_{2}\sigma _2^2}}}$.
}

  \emph{Proof:} Please see the proof in the Appendix \ref{app_lemma_3}.\qed

Then, let $R_{{\rm{LiFi - WiFi}}}^{\rm{L}}$ and $R_{{\rm{LiFi - WiFi}}}^{\rm{U}}$  respectively denote the       lower bound and upper bound of  ${R_{{\rm{LiFi - WiFi}}}}$, which are given by
\begin{subequations}
\begin{align}
 R_{{\rm{LiFi - WiFi}}}^{\rm{L}}  \triangleq &{R^{\rm{L}}_{{\rm{LiFi}}}}+ {R^{\rm{L}}_{{\rm{WiFi}}}}.\label{rate_up_low}\\
 R_{{\rm{LiFi - WiFi}}}^{\rm{U}} \triangleq& {R^{\rm{U}}_{{\rm{LiFi}}}}+ {R^{\rm{U}}_{{\rm{WiFi}}}}.\label{rate_up_up}
\end{align}
  \end{subequations}
{   The lower bound $R_{\mathrm{LiFi}}^{\mathrm{L}}$, $R_{\mathrm{WiFi}}^{\mathrm{L}}$ $R_{\mathrm{LiFi-WiFi}}^{\mathrm{L}}$ are not tight, and might be negative \cite{Zeng2012WCL}. When $P_{\mathrm{T}}\to 0$ and $P_{\mathrm{T}}\to +\infty$, there is a constant gap, i.e., $B_1\left(\frac{1}{\ln 2} - 1\right)$, between the $R_{\mathrm{LiFi}}$ \eqref{rate_LiFi} and $R_{\mathrm{LiFi}}^{\mathrm{L}}$ \eqref{rate_LiFi_low}. There also is a constant gap, i.e., $B_2\left(\frac{1}{\ln 2} - 1\right)$, between the $R_{\mathrm{WiFi}}$ \eqref{rate_WiFi} and $R_{\mathrm{WiFi}}^{\mathrm{L}}$ \eqref{rate_WiFi_L1_1}. Thus, when $P_{\mathrm{T}}\to 0$ and $P_{\mathrm{T}}\to +\infty$, there is a constant gap, i.e., $\left(\frac{1}{\ln 2} - 1\right)\left(B_1 + B_2\right)$, between the $R_{\mathrm{LiFi-WiFi}}$ \eqref{rate_LiFi_WiFi} and $R_{\mathrm{LiFi-WiFi}}^{\mathrm{L}}$ \eqref{rate_up_low}. If the constant gap is directly added to $R_{\mathrm{LiFi-WiFi}}^{\mathrm{L}}$, there might be intersections between $R_{\mathrm{LiFi-WiFi}}$ and $R_{\mathrm{LiFi-WiFi}}^{\mathrm{L}}$ as similar as Fig. 1-3 in \cite{Zeng2012WCL}.}

Based on the lower bound  of the achievable rate $R_{{\rm{LiFi - WiFi}}}^{\rm{L}}$, we aim to obtain the optimal  distribution of the signal magnitudes and power allocation jointly, which
  can be
formulated as
\begin{align}\label{max_rate_agg}
 \mathop {\max }\limits_{q_1, q_2,\left\{ {{p_{1,k}}} \right\},\left\{ {{p_{2,l}}} \right\}}  \quad & R_{{\rm{LiFi - WiFi}}}^{\rm{L}}\\
 {\rm{s.t. }} \quad\quad\quad & \eqref{max_rate_accuracya}, \eqref{max_rate_accuracyb}, \eqref{max_rate_accuracyc},  \eqref{max_rate_accuracyd}, \eqref{max_rate_accuracye}, \eqref{max_rate_accuracyf} \nonumber.
\end{align}

Furthermore, by defining   ${{{\hat q}_1}} \buildrel \Delta \over = q_1^2$ and ${{{\hat q}_2}} \buildrel \Delta \over = {\left| {{q_2}} \right|^2}$, the lower bound of the achievable rate $R_{{\rm{LiFi - WiFi}}}^{\rm{L}}$ can be  rewritten as
\begin{align}\label{rate_agg_hat}
R_{{\rm{LiFi - WiFi}}}^{\rm{L}}=  &\left( {1 - \frac{1}
{{\ln 2}}} \right)\left({{{B_1} + {B_2}}}\right)\nonumber\\
&~~~~~-2{B_1}{\bf{p}}_1^{\rm{T}}{\bf{u}}\left( {{{\hat q}_1}} \right)- {B_2}{\bf{p}}_2^{\rm{T}}{\bf{v}}\left( {{{\hat q}_2}} \right),
\end{align}
where\begin{subequations}\label{definition}
\begin{align}
&{\bf{u}}\left( {{\hat q_1}} \right) \buildrel \Delta \over= {\left[ {{\log _2}{\bf{a}}_1^{\rm{T}}\left( {{\hat q_1}} \right){{\bf{p}}_1}, \ldots ,{\log _2}{\bf{a}}_M^{\rm{T}}\left( {{\hat q_1}} \right){{\bf{p}}_1}} \right]^{\rm{T}}},\\
&{{\bf{a}}_k}\left( {{\hat q_1}} \right) \buildrel \Delta \over= {\left[ {{a_{k,1}}\left( {{\hat q_1}} \right), \ldots ,{a_{k,M}}\left( {{\hat q_1}} \right)} \right]^{\rm{T}}}, \\
&{{a_{k,m}}\left( {{{\hat q}_1}} \right)} \buildrel \Delta \over =\exp \left( { - \frac{{{{\left( {{x_{1,k}} - {x_{1,m}}} \right)}^2}{g_1^2}{{{\hat q}_1}}}}{{4B_{1}\sigma _1^2}}} \right), \forall {k,m} \in {\cal M},\\
&{\bf{v}}\left( {{\hat q_2}} \right) \buildrel \Delta \over= {\left[ {{\log _2}{\bf{b}}_1^{\rm{T}}\left( {{\hat q_2}} \right){{\bf{p}}_2}, \ldots ,{\log _2}{\bf{b}}_N^{\rm{T}}\left( {{\hat q_2}} \right){{\bf{p}}_2}} \right]^{\rm{T}}},\\
&{{\bf{b}}_l}\left( {{\hat q_2}} \right) \buildrel \Delta \over= {\left[ {{b_{l,1}}\left( {{\hat q_2}} \right), \ldots ,{b_{l,N}}\left( {{\hat q_2}} \right)} \right]^{\rm{T}}}, \\
&{{b_{l,n}}\left( {{{\hat q}_2}} \right)} \buildrel \Delta \over = \exp \left( { - \frac{{{{\left| {{x_{2,l}} - {x_{2,n}}} \right|}^2}{{\left| {{g_2}} \right|}^2}{{{\hat q}_2}}}}{{2B_{2}\sigma _2^2}}} \right),  \forall {l,n} \in {\cal N}.
\end{align}
\end{subequations}

Then,  problem \eqref{max_rate_agg}  can be
formulated as
\begin{align}\label{max_rate_agg_1}
 \mathop {\min }\limits_{{{{\hat q}_1}}, {{{\hat q}_2}},{{\bf{p}}_1 },{{\bf{p}}_2}} \quad &  2{B_1}{\bf{p}}_1^{\rm{T}}{\bf{u}}\left( {{{\hat q}_1}} \right)+ {B_2}{\bf{p}}_2^{\rm{T}}{\bf{v}}\left( {{{\hat q}_2}} \right) \\
 {\rm{s}}{\rm{.t}}{\rm{.  }} \quad\quad & {\eqref{max_rate_accuracy_2b}, \eqref{max_rate_accuracy_2c}, \eqref{max_rate_accuracy_2d}.}\nonumber
\end{align}

It can be easily checked that  the objection function \eqref{max_rate_agg_1}  is convex with respect to ${{{\hat q}_1}}$ and ${{{\hat q}_2}}$, and constraints \eqref{max_rate_accuracy_2b}, \eqref{max_rate_accuracy_2c}, and \eqref{max_rate_accuracy_2d} are linear in terms of ${{{\hat q}_1}}$, and ${{{\hat q}_2}}$. Therefore, problem \eqref{max_rate_agg_1} is convex   with respect to (w.r.t.) ${{{\hat q}_1}}$ and ${{{\hat q}_2}}$.
While, the problem   \eqref{max_rate_agg_1}  is nonconvex w.r.t. ${\bf{p}}_1$ and ${\bf{p}}_2$.
To solve problem \eqref{max_rate_agg_1}, we  by    solving the following two sub-problems iteratively, and until the objective function converges.

\subsection{\textbf{Power Allocation Sub-problem}}

Firstly, we  solve the power allocation sub-problem:  optimizing the LiFi and the WiFi links power allocation, i.e., $ {{\hat { q}_1}}$ and $ {{\hat { q}_2}}$, with fixed ${\bf{p}}_1$ and ${\bf{p}}_2$.
Then,
  problem \eqref{max_rate_agg_1} is reformulated as
\begin{subequations}\label{max_rate_W}
\begin{align}
 \mathop {\min }\limits_{{{{\hat q}_1}},{{{\hat q}_2}}} \quad &2 {B_1}{\bf{p}}_1^{\rm{T}}{\bf{u}}\left( {{{\hat q}_1}} \right)+ {B_2}{\bf{p}}_2^{\rm{T}}{\bf{v}}\left( {{{\hat q}_2}} \right)  \\
 {\rm{s}}{\rm{.t}}{\rm{.  }} \quad &  {\eta_1}P_{\mathrm{e},1}{{{\hat q}_1}} + {\eta_2}P_{\mathrm{e},2}{{{\hat q}_2}} \le {P_{\rm{T}}},\label{max_rate_W_b}\\
 & \hat q_1 \le \tau^2, \label{max_rate_W_op}\\
 & {{{\hat q}_1}} \ge 0, {{{\hat q}_2}} \ge 0.\label{max_rate_W_c}
\end{align}
\end{subequations}

Since  ${\log _2} x $ is a concave
 function,    at least one of   constraint \eqref{max_rate_W_b}   and constraint \eqref{max_rate_W_op}  is active for the optimal power allocation, i.e., $ {\eta_1}P_{\mathrm{e},1}{{{\hat q}_1}} + {\eta_2}P_{\mathrm{e},2}{{{\hat q}_2}} = {P_{\rm{T}}}$ or $\hat q_1 = \tau^2$.
 When $\hat q_1 = \tau^2$,   constraint \eqref{max_rate_W_b} is also active.
  Therefore,  constraint \eqref{max_rate_W_b}
 is always active, i.e., $ {\eta_1}P_{\mathrm{e},1}{{{\hat q}_1}} + {\eta_2}P_{\mathrm{e},2}{{{\hat q}_2}} = {P_{\rm{T}}}$.
In the following, we will discuss the optimal power allocation of    problem \eqref{max_rate_W} based on   whether constraint \eqref{max_rate_W_op} is active or not.

\subsubsection{${\eta_1}{\varepsilon _1} \tau^2 >{P_{\rm{T}}}$}

In this case, constraint \eqref{max_rate_W_op} is inactive, i.e., $\hat q_1 < \tau^2$.
Problem \eqref{max_rate_W} can be   reformulated as
\begin{subequations}\label{max_rate_W_inac}
\begin{align}
 \mathop {\min }\limits_{{{{\hat q}_1}},{{{\hat q}_2}}} \quad & 2{B_1}{\bf{p}}_1^{\rm{T}}{\bf{u}}\left( {{{\hat q}_1}} \right)+ {B_2}{\bf{p}}_2^{\rm{T}}{\bf{v}}\left( {{{\hat q}_2}} \right) \label{max_rate_W_a}\\
 {\rm{s}}{\rm{.t}}{\rm{.  }}\quad & {\eta_1}P_{\mathrm{e},1}{{{\hat q}_1}} + {\eta_2}P_{\mathrm{e},1}{{{\hat q}_2}} = {P_{\rm{T}}},\\
 & {{{\hat q}_1}} \ge 0, {{{\hat q}_2}} \ge 0.\label{max_rate_W_c}
\end{align}
\end{subequations}
Substituting ${{{\hat q}_2}} = \frac{{{P_{\rm{T}}} - {\eta_1}{\varepsilon _1}{{{\hat q}_1}}}}{{{\eta_2}{\varepsilon _2}}}$ into problem \eqref{max_rate_W_inac}, the objective function  \eqref{max_rate_W_a} is reformulated  as
\begin{align}
\Phi \left({{\hat q}_1}\right)  \buildrel \Delta \over =  2{B_1}{\bf{p}}_1^{\rm{T}}{\bf{u}}\left( {{{\hat q}_1}} \right)+ {B_2}{\bf{p}}_2^{\rm{T}}{\bf{\hat v}}\left( {{{\hat q}_1}} \right),
\end{align}
where ${\bf{\hat v}}\left( {{\hat q_1}} \right) \buildrel \Delta \over = {\left[ { {\log _2}{\bf{\hat b}}_1^{\rm{T}}\left( {{\hat q_1}} \right){{\bf{p}}_1}, \ldots , {\log _2}{\bf{\hat b}}_N^{\rm{T}}\left( {{\hat q_1}} \right){{\bf{p}}_1}} \right]^{\rm{T}}}$, ${{\bf{\hat b}}_l}\left( {{\hat q_1}} \right) \buildrel \Delta \over = {\left[ {{\hat b_{l,1}}\left( {{\hat q_1}} \right), \ldots ,{\hat b_{l,N}}\left( {{\hat q_1}} \right)} \right]^{\rm{T}}}, \forall {l} \in {\cal N}$, and ${{\hat b}_{l,n}}\left( {{{\hat q}_1}} \right)\buildrel \Delta \over = \exp \left( { - \frac{{{{\left| {{x_{2,l}} - {x_{2,n}}} \right|}^2}{{\left| {{g_2}} \right|}^2}\left( {{P_{\rm{T}}} - {\eta _1}{\varepsilon _1}{{\hat q}_1}} \right)}}{{2{B_2}\sigma _2^2{\eta _2}{\varepsilon _2}}}} \right), \forall {l,n} \in {\cal N}$.

Thus, problem \eqref{max_rate_W} can be reformulated as
\begin{subequations}\label{max_rate_W1}
\begin{align}
 \mathop {\min }\limits_{{{{\hat q}_1}}}  & \quad \Phi \left({{\hat q}_1}\right)\\
 {\rm{s}}{\rm{.t}}{\rm{.  }} & \quad 0 \le {{{\hat q}_1}}  \le \frac{P_{\rm{T}}}{{{{\eta_1}{\varepsilon _1}}}}.
\end{align}
\end{subequations}

Let ${{\hat q}_1}^{{\rm{opt}}}$ denote the optimal solution of problem \eqref{max_rate_W1},
 and    $\Phi ^{{\rm{opt}}}$ denote the maximum rate of the  aggregated LiFi - WiFi  system.
Then, we can obtain a stationary point ${{\hat q}_1}^{\rm{sta}}$  by setting $\frac{{\partial \Phi \left({{\hat q}_1}\right) }}{{\partial {{{\hat q}_1}}}}=0$
 so that the optimal solution ${{\hat q}_1}^{\rm{sta}}$ satisfies the following equation
 \begin{align}\label{sta_W1}
2{B_1}{\bf{p}}_1^{\rm{T}}{\bf{\tilde a}}\left( {{{\hat q}_1}} \right) = {B_2}{\bf{p}}_2^{\rm{T}}{{{\bf{\tilde b}}}}\left( {{{\hat q}_1}} \right),
\end{align}
where \begin{subequations}\label{definition_2}
\begin{align}
&{\bf{\tilde a}}\left( {{{\hat q}_1}} \right) = {\left[ {{{\tilde a}_1}\left( {{{\hat q}_1}} \right), \ldots ,{{\tilde a}_M}\left( {{{\hat q}_1}} \right)} \right]^{\rm{T}}},\\
&{{\tilde a}_k}\left( {{{\hat q}_1}} \right) = \frac{{{{ {{\bf{c}}^{\rm{T}}_k \odot {{\bf{a}}^{\rm{T}}_k}\left( {{{\hat q}_1}} \right)} }}{{\bf{p}}_1}}}{{{{\bf{a}}^{\rm{T}}_k}\left( {{{\hat q}_1}} \right){{\bf{p}}_1}}}, \\
&{{\bf{c}}_k} \buildrel \Delta \over = {\frac{{g_1^2}}{{4{B_1}\sigma _1^2}}} {\left[ {{{\left( {{x_{1,k}} - {x_{1,1}}} \right)}^2}, \ldots ,{{\left( {{x_{1,k}} - {x_{1,M}}} \right)}^2}  } \right]^{\rm{T}}},\forall {k} \in {\cal M}, \\
&{\bf{\tilde b}}\left( {{{\hat q}_1}} \right) = {\left[ {{{\tilde b}_1}\left( {{{\hat q}_1}} \right), \ldots ,{{\tilde b}_N}\left( {{{\hat q}_1}} \right)} \right]^{\rm{T}}},\\
&{{\tilde b}_l}\left( {{{\hat q}_1}} \right) = \frac{{{{ {{\bf{d}}^{\rm{T}}_l \odot {{{\bf{\hat b}}}^{\rm{T}}_l}\left( {{{\hat q}_1}} \right)} }}{{\bf{p}}_2}}}{{{{{\bf{\hat b}}}^{\rm{T}}_l}\left( {{{\hat q}_1}} \right){{\bf{p}}_2}}}, \\
&{{\bf{d}}_l}\buildrel \Delta \over ={\frac{{{\eta _1}{\varepsilon _1}{{\left| {{g_2}} \right|}^2}}}{{2{B_2}\sigma _2^2{\eta _2}{\varepsilon _2}}}} {\left[ {{{\left| {{x_{2,l}} - {x_{2,1}}} \right|}^2}, \ldots ,{{\left| {{x_{2,l}} - {x_{2,N}}} \right|}^2}} \right]^{\rm{T}}}, \forall {l} \in {\cal N}.
\end{align}
\end{subequations}

Let $\Phi ^{{\rm{opt},1}}$    and ${{\hat q}_1}^{{\rm{opt},1}}$  respectively denote   the minimal objective function and the optimal power allocation ${{\hat q}_1}$ of  problem \eqref{max_rate_W}.
 Thus, if $0 \le {{\hat q}_1}  \le \frac{P_{\rm{T}}}{{{{\eta_1}{\varepsilon _1}}}}$, we have
\begin{subequations}
\begin{align}
&\Phi ^{{\rm{opt},1}} = \min \left\{ {\Phi \left( {{{{\hat q}_1}^{\rm{sta}}}} \right),\Phi \left( 0 \right),\Phi \left( \frac{P_{\rm{T}}}{{{{\eta_1}{\varepsilon _1}}}} \right)} \right\},\\
&{{\hat q}_1}^{{\rm{opt},1}}= \mathop {\arg \min }\limits_{{{\hat q}_1}} \left\{ {\Phi \left( {{{{\hat q}_1}^{\rm{sta}}}} \right),\Phi \left( 0 \right),\Phi \left( \frac{P_{\rm{T}}}{{{{\eta_1}{\varepsilon _1}}}} \right)} \right\}.
\end{align}
\end{subequations}
Otherwise, we have
\begin{subequations}
\begin{align}
&\Phi ^{{\rm{opt},1}} = \min \left\{ {\Phi \left( 0 \right),\Phi \left( \frac{P_{\rm{T}}}{{{{\eta_1}{\varepsilon _1}}}} \right)} \right\},\\
&{{\hat q}_1}^{{\rm{opt},1}} =\mathop {\arg \min }\limits_{{{\hat q}_1}} \left\{ {\Phi \left( 0 \right),\Phi \left( \frac{P_{\rm{T}}}{{{{\eta_1}{\varepsilon _1}}}} \right)} \right\}.
\end{align}
\end{subequations}

\subsubsection{${\eta_1}{{P_{e,1}}} \tau^2 \le {P_{\rm{T}}}$}
When ${\eta_1}{{P_{e,1}}} \tau^2 \le {P_{\rm{T}}}$, constraint \eqref{max_rate_W_op} could be either active or inactive, i.e., $\hat q_1 = \tau^2$ or $\hat q_1 < \tau^2$.
 When constraint \eqref{max_rate_W_op} is inactive, the minimal objective function of  problem \eqref{max_rate_W} is the same as the case ${\eta_1}{{P_{e,1}}} \tau^2 < {P_{\rm{T}}}$.
Otherwise, the optimal allocated power of problem \eqref{max_rate_W} are  $\hat q_1 = \tau^2 $, and ${{{\hat q}_2}} = \frac{{{P_{\rm{T}}} - {\eta_1}{{P_{e,1}}}\tau^2}}{{{\eta_2}{{P_{e,2}}}}}$. Here, the corresponding  minimal objective function of  problem \eqref{max_rate_W} is    $\Phi \left( \tau^2 \right)$.
Let  $\Phi ^{{\rm{opt},2}}$ and ${{\hat q}_1}^{{\rm{opt},2}}$ respectively denote the minimal objective function   and    the optimal power allocation ${{\hat q}_1}$ of  problem \eqref{max_rate_W}, which are given by
\begin{subequations}
\begin{align}
&\Phi ^{{\rm{opt},2}} = \min \left\{ {\Phi ^{{\rm{opt},1}},\Phi \left( \tau^2 \right)} \right\},\\
&{{\hat q}_1}^{{\rm{opt},2}} =\mathop {\arg \min }\limits_{{{\hat q}_1}} \left\{ {\Phi ^{{\rm{opt},1}},\Phi \left( \tau^2 \right)}  \right\}.
\end{align}
\end{subequations}

By solving the power allocation sub-problem, we can obtain the  optimal  power of LiFi and WiFi links $q_1$ and $q_2$.

\subsection{\textbf{Probability Distribution Sub-problem}}
Next, we  solve the probability distribution sub-problem:  optimizing the LiFi and the WiFi links probability distribution, i.e., ${\bf{p}}_1$ and ${\bf{p}}_2$, with  given $ {{\hat { q}_1}}$ and $ {{\hat { q}_2}}$.
Then,  the    maximization
problem of  the aggregated LiFi-WiFi system  \eqref{max_rate_W}  can be
formulated as
\begin{subequations}\label{max_rate_3}
\begin{align}
 \mathop {\min }\limits_{{{\bf{p}}_1},{{\bf{p}}_2}}   \quad &2{B_1}{\bf{p}}_1^{\rm{T}}{\bf{u}}\left( {\bf{p}}_1 \right) +{B_2}{\bf{p}}_2^{\rm{T}}{\bf{v}}\left( {\bf{p}}_2 \right) \\
 {\rm{s}}{\rm{.t}}{\rm{.  }} \quad &{{\bf{p}}_1 \in {\Upsilon _1}}, ~{{\bf{p}}_2 \in {\Upsilon _2}},
\end{align}
\end{subequations}
where ${\bf{u}}\left( {{{\bf{p}}_1}} \right) = {\left[ {{\log _2}{{\bf{a}}^{\rm{T}}_1}\left( {{{\hat q}_1}} \right){{\bf{p}}_1}, \ldots ,{\log _2}{{\bf{a}}^{\rm{T}}_M}\left( {{{\hat q}_1}} \right){{\bf{p}}_1}} \right]^{\rm{T}}}$, and ${\bf{v}}\left( {{{\bf{p}}_2}} \right) = {\left[ {{\log _2}{{\bf{b}}^{\rm{T}}_1}\left( {{{\hat q}_2}} \right){{\bf{p}}_2}, \ldots ,{\log _2}{{\bf{b}}^{\rm{T}}_N}\left( {{{\hat q}_2}} \right){{\bf{p}}_2}} \right]^{\rm{T}}}$.
 Then, problem \eqref{max_rate_3}  can be divided into two independent subproblems as
\begin{subequations}\label{max_rate_LiFi}
\begin{align}
 \mathop {\min }\limits_{{{\bf{p}}_1 }}\quad & f_1\left( {{{{\bf{ p}}}_{1}}} \right)\\
  {\rm{s}}{\rm{.t}}{\rm{.  }}\quad &{{\bf{p}}_1 \in {\Upsilon _1}},
\end{align}
\end{subequations}
and
\begin{subequations}\label{max_rate_WiFi}
\begin{align}
 \mathop {\min }\limits_{{{\bf{p}}_2 }}\quad & f_2\left( {{{{\bf{ p}}}_{2}}} \right)\\
  {\rm{s}}{\rm{.t}}{\rm{.  }} \quad&{{\bf{p}}_2 \in {\Upsilon _2}}.
\end{align}
\end{subequations}
where $f_1\left( {{{{\bf{ p}}}_{1}}} \right) \buildrel \Delta \over =   2{B_1}{\bf{p}}_1^{\rm{T}}{\bf{u}}\left( {\bf{p}}_1 \right)$, and
$f_2\left( {{{{\bf{ p}}}_{2}}} \right)  \buildrel \Delta \over =    {B_2}{\bf{p}}_2^{\rm{T}}{\bf{v}}\left( {\bf{p}}_2 \right)$.
We can adopt the Frank-Wolfe method \cite{Dimitri} to solve
problem \eqref{max_rate_LiFi} and \eqref{max_rate_WiFi}.
Specifically, let ${{\bf{p}}^{\left[ {i} \right]}_{1}}$ denote a feasible point at  the $i$th  iteration.
The first-order Taylor expansion of $f_1\left( {{{{\bf{ p}}}_{1}}} \right)$ is given by
\begin{align}\label{frank}
{f_1}\left( {{{\bf{p}}_1}} \right) \approx {f_1}\left( {{{\bf{p}}^{\left[ {i} \right]}_{1}}} \right) + {\nabla _{{{\bf{p}}_1}}}f_{\rm{1}}^{\rm{T}}\left( {{{\bf{p}}^{\left[ {i} \right]}_{1}}} \right)\left( {{{\bf{p}}_1} - {{\bf{p}}^{\left[ {i} \right]}_{1}}} \right),
\end{align}
where ${\nabla _{{{\bf{p}}_1}}}f_{\rm{1}}^{\rm{T}}\left( {{{\bf{p}}^{\left[ {i} \right]}_{1}}} \right)$
denotes the gradient of the objective function $f_1\left( {{{{\bf{ p}}}_{1}}} \right) $.
 Since ${{\bf{p}}^{\left[ {i} \right]}_{1}}$ is obtained, problem \eqref{max_rate_LiFi}  is equivalent to
 \begin{subequations}\label{frank_app}
\begin{align}
 \mathop {\min }\limits_{{{\bf{p}}_{1} }}\quad &   {\nabla _{{{\bf{p}}_1}}}f_{\rm{1}}^{\rm{T}}\left( {{{\bf{p}}^{\left[ {i} \right]}_{1}}} \right){{\bf{p}}_1}\\
  {\rm{s}}{\rm{.t}}{\rm{.  }}\quad &{{\bf{p}}_1 \in {\Upsilon _1}}.
\end{align}
\end{subequations}

Then, the $i+1$th  iteration point $ {{{\bf{p}}^{\left[ {i+1} \right]}_{1}}}$ is updated by
 \begin{align}
 {{{\bf{p}}^{\left[ {i+1} \right]}_{1}}} ={{{\bf{p}}^{\left[ {i} \right]}_{1}} + {\lambda^{\left[ {i} \right]}_{1}}{{\bf{d}}^{\left[ {i} \right]}_{1}}},
 \end{align}
 where ${\lambda ^{\left[ {i} \right]}_{1}} $ denotes the stepsize of the $i$th iteration, and ${{\bf{d}}^{\left[ {i} \right]}_{1}}$ denotes the feasible descending direction of the $i$th iteration.
The details of the  Frank-Wolfe method   are summarized   in Algorithm 4, which  outputs the probability distribution of the LiFi  link, i.e., ${\bf{p}}_1$.
Similar as problem  \eqref{max_rate_LiFi}, we can obtain the optimal   probability distribution of  the WiFi  link, i.e., ${\bf{p}}_2$,  by solving problem \eqref{max_rate_WiFi} through  Frank-Wolfe method. We
omit the detailed derivations for brevity. {The optimal  probability distribution $\mathbf{p}_1$, \(\mathbf{p}_2\) can be achieved by the distribution matching (DM) \cite{Schulte2016,Chen2017,Buchali2016}.}

In summary, the achievable rate maximization of the aggregated LiFi-WiFi system \eqref{max_rate_agg_1} can
be solved by Algorithm 5. The optimal power allocation of  LiFi and WiFi links $ {{\hat { q}_1}}$ and $ {{\hat { q}_2}}$, the  probability distribution of LiFi and WiFi links ${\bf{p}}_1$ and ${\bf{p}}_2$, and the maximal lower bound of the achievable rate ${\hat R}_{\rm{LiFi-WiFi}}^{\rm{L}}$  can be obtained by Algorithm 3. Note that, based on the upper bound  of the achievable rate  $R_{{\rm{LiFi - WiFi}}}^{\rm{U}}$, we can also find the optimal   distribution of  the  signal magnitudes and power allocation to   maximize  the  achievable rate
of  the aggregated LiFi-WiFi system, which is similar to the lower bound case.

\begin{algorithm}[htbp]
    \caption{  Frank-Wolfe Method for the Probability Distribution Sub-Problem \eqref{max_rate_3}.}
    \label{alg:Framwork_L_1}
    \begin{algorithmic}[1]
        \State \textbf{Initialization}: Given  $\varsigma \ge 0$, set $i=1$, and choose ${{{{\bf{p}}}^{\left[ {1} \right]}_{1}}} \in {\Upsilon _1}$.
        \State \textbf{repeat}
        \State \quad $i \leftarrow i+1$.
        \State \quad Obtain ${{{\bf{\hat p}}}^{\left[ {i-1} \right]}_{1}}$ by solving  problem  \eqref{frank_app}.
        \State  \quad Construct feasible descending direction ${{\bf{d}}^{\left[ {i-1} \right]}_{1}} = {{{\bf{\hat p}}}^{\left[ {i-1} \right]}_{1}}-{{{{\bf{p}}}^{\left[ {i-1} \right]}_{1}}}$.
        \State \quad Use the bisection search to obtain the optimal ${\lambda ^{\left[ {i-1} \right]}_{1}}=\mathop {\arg \min }\limits_{{\lambda _1}} {f_1}\left( {{\bf{p}}_1^{\left[ {i - 1} \right]} + {\lambda _1}{\bf{d}}_1^{\left[ {i - 1} \right]}} \right)$.
        \State \quad Update ${{{\bf{p}}^{\left[ {i} \right]}_{1}}} ={{{\bf{p}}^{\left[ {i-1} \right]}_{1}} + {\lambda^{\left[ {i-1} \right]}_{1}}{{\bf{d}}^{\left[ {i-1} \right]}_{1}}}$.
        \State \textbf{until} $\left| {\nabla _{{{\bf{p}}_1}}}f_{\rm{1}}^{\rm{T}}\left( {{{\bf{p}}^{\left[ {i-1} \right]}_{1}}} \right) {{\bf{d}}^{\left[ {i-1} \right]}_{1}} \right| \le \varsigma $.
        \State
        \Return ${{{\bf{p}}_{1}}} ={{{\bf{p}}^{\left[ {i+1} \right]}_{1}}} $.
    \end{algorithmic}
\end{algorithm}

\begin{algorithm}[htbp]
    \caption{ Optimal discrete constellation input distributions based on    lower bounds }
    \label{alg:Framwork_L_2}
    \begin{algorithmic}[1]
        \State \textbf{Initialization}: Given  $\xi  \ge 0$, set $k=1$, and choose ${{{{\bf{p}}}_{1}^{\left[ 1 \right]}}} \in {\Upsilon _1}$, ${{{{\bf{p}}}_{2}^{\left[ 1 \right]}}} \in {\Upsilon _2}$.
        \State  Update ${{\hat q}_{1}^{\left[ 1 \right]}}$ and ${{\hat q}_{2}^{\left[ 1 \right]}}$  by solving problem \eqref{max_rate_W}.
        \State Obtain ${ R}^{{\rm{L}}\left[ 1\right]}_{\rm{LiFi-WiFi}}$ by substituting ${\hat q}_{1}^{\left[ 1  \right]}$, ${\hat q}_{2}^{\left[ 1  \right]}$, ${{\bf{ p}}_{1}^{\left[ 1 \right]}}$ and  ${{\bf{ p}}_{2}^{\left[ 1  \right]}}$  into formulation \eqref{rate_agg_hat}.
        \State \textbf{repeat}.
        \State \quad $k \leftarrow k+1$.
        \State  \quad Update ${\hat q}_{1}^{\left[ k  \right]}$ and ${\hat q}_{2}^{\left[ k  \right]}$ by solving problem \eqref{max_rate_W}.
        \State \quad Update ${{\bf{ p}}_{1}^{\left[ k \right]}}$ and  ${{\bf{ p}}_{2}^{\left[ k  \right]}}$ by Frank-Wolfe method (in Algorithm 4).
        \State \quad Obtain ${{ R}_{\rm{LiFi-WiFi}}^{{\rm{L}}{\left[ k \right]}}}$ by substituting
        ${\hat q}_{1}^{\left[ k  \right]}$, ${\hat q}_{2}^{\left[ k  \right]}$, ${{\bf{ p}}_{1}^{\left[ k \right]}}$ and  ${{\bf{ p}}_{2}^{\left[ k  \right]}}$ into formulation \eqref{rate_agg_hat}.
        \State \textbf{ until} $\left| { R}^{{\rm{L}}\left[ k  \right]}_{\rm{LiFi-WiFi}} - { R}^{{\rm{L}}\left[ k -1\right]}_{\rm{LiFi-WiFi}} \right| \le \xi $.
        \State
        \Return ${{{\bf{p}}_{1}}} ={{{\bf{p}}^{\left[ k \right]}_{1}}} $, ${{{\bf{p}}_{2}}} ={{{\bf{p}}^{\left[ k \right]}_{2}}} $, ${\hat q}_1= {\hat q}^{\left[ k \right]}_1$ and ${\hat q}_2= {\hat q}^{\left[ k \right]}_2$,${ R}_{\rm{LiFi-WiFi}}^{\rm{L}}={ R}_{\rm{LiFi-WiFi}}^{{\rm{L}}\left[ k \right]}$ .
    \end{algorithmic}
\end{algorithm}

\section{Simulation Results and Discussions}

In this section, we illustrate the performance  of the proposed schemes for the considered aggregated
LiFi-WiFi  system.
Moreover, the simulation results also   demonstrate the impact of key parameters    on  both the achievable rate and energy efficiency
  of the aggregated LiFi-WiFi   system, such as total power threshold, bandwidths,  etc.
In the simulations, we consider the locations of the LED and PD as $\left(0,0,5.7\right)$m and $\left(0,0,1.7\right)$m, respectively.
More detailed system  parameters of the LiFi and WiFi links are
given in Table I.
\begin{table}[htbp]
    \centering
    \caption{\normalsize Basic parameters }
    \begin{tabular}{|l|l|l|l|}
        \hline
        \multicolumn{2}{|c|}{LiFi Link}&\multicolumn{2}{c|}{WiFi Link}\\
        \hline
        Parameters  & Value  &Parameters  & Value \\
        \hline
        ${\sigma_1 ^2}$ & $10^{-21}  \rm{A^2/Hz}$ & ${\sigma_2 ^2}$ &  -57dBm/MHz\\
        \hline
        FoV~${\Psi _c}$ & ${90 ^ \circ }$ & $d_B$ & 5 m \\
        \hline
        $A_{\rm{PD}}$ & $1 {\rm{c}}{{\rm{m}}^2}$ &  $d_{2}$ & 4 m   \\
        \hline
        $I_{\rm{H}}$ & 8 A &   $f_c$ & 2.4 GHz \\
        \hline
        ${\theta _{1/2}}$ & ${60 ^ \circ }$ & $\psi$ & ${45 ^ \circ }$ \\
        \hline
        $B_1$ &$40$MHz &{ $B_2$}&{ $20$MHz}\\
        \hline
    \end{tabular}
\end{table}

\subsection{Performance    of the Aggregated LiFi-WiFi System based on   ${{{R}}_{{\rm{LiFi-WiFi}}}}$ \eqref{rate_LiFi_WiFi}}

\begin{figure}[htbp]
    \begin{minipage}[b]{0.45\textwidth}
        \centering
        \includegraphics[height=5cm]{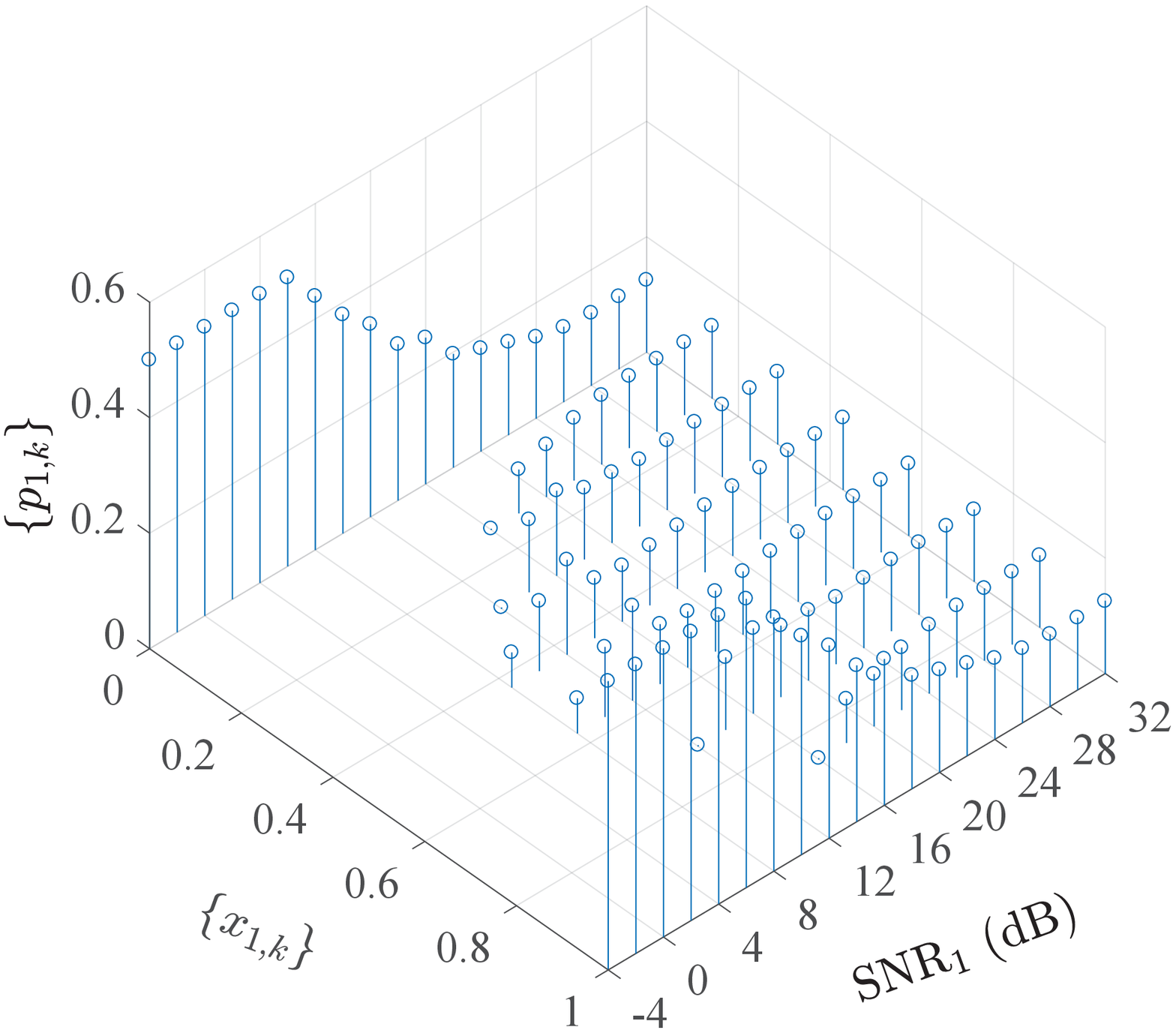}
        \vskip-0.2cm\centering {\footnotesize (a)}
    \end{minipage}\hfill

\begin{minipage}[b]{0.45\textwidth}
    \centering
    \includegraphics[height=5cm]{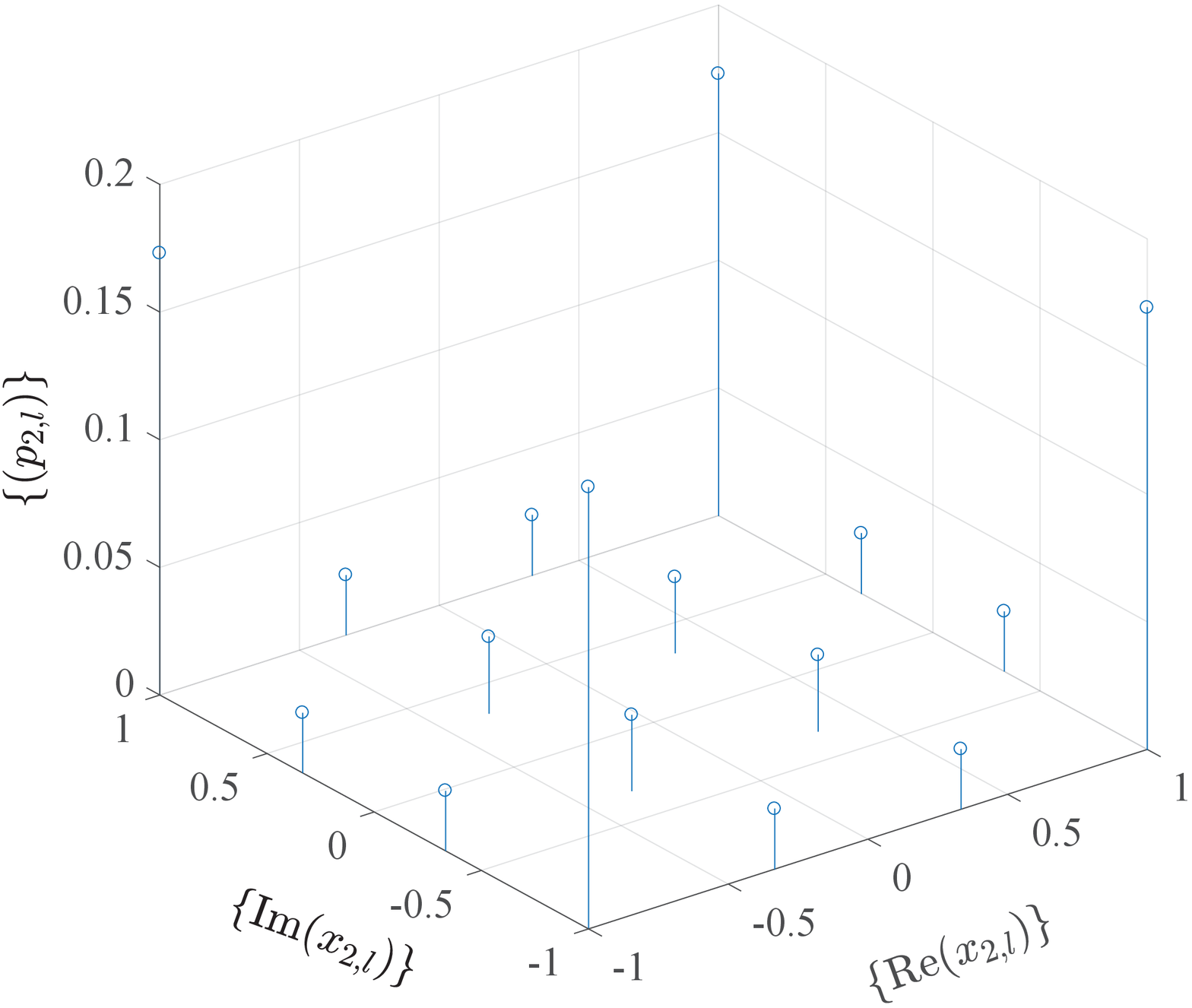}
    \vskip-0.2cm\centering {\footnotesize (b)}
\end{minipage}\hfill

    \begin{minipage}[b]{0.45\textwidth}
        \centering
        \includegraphics[height=5cm]{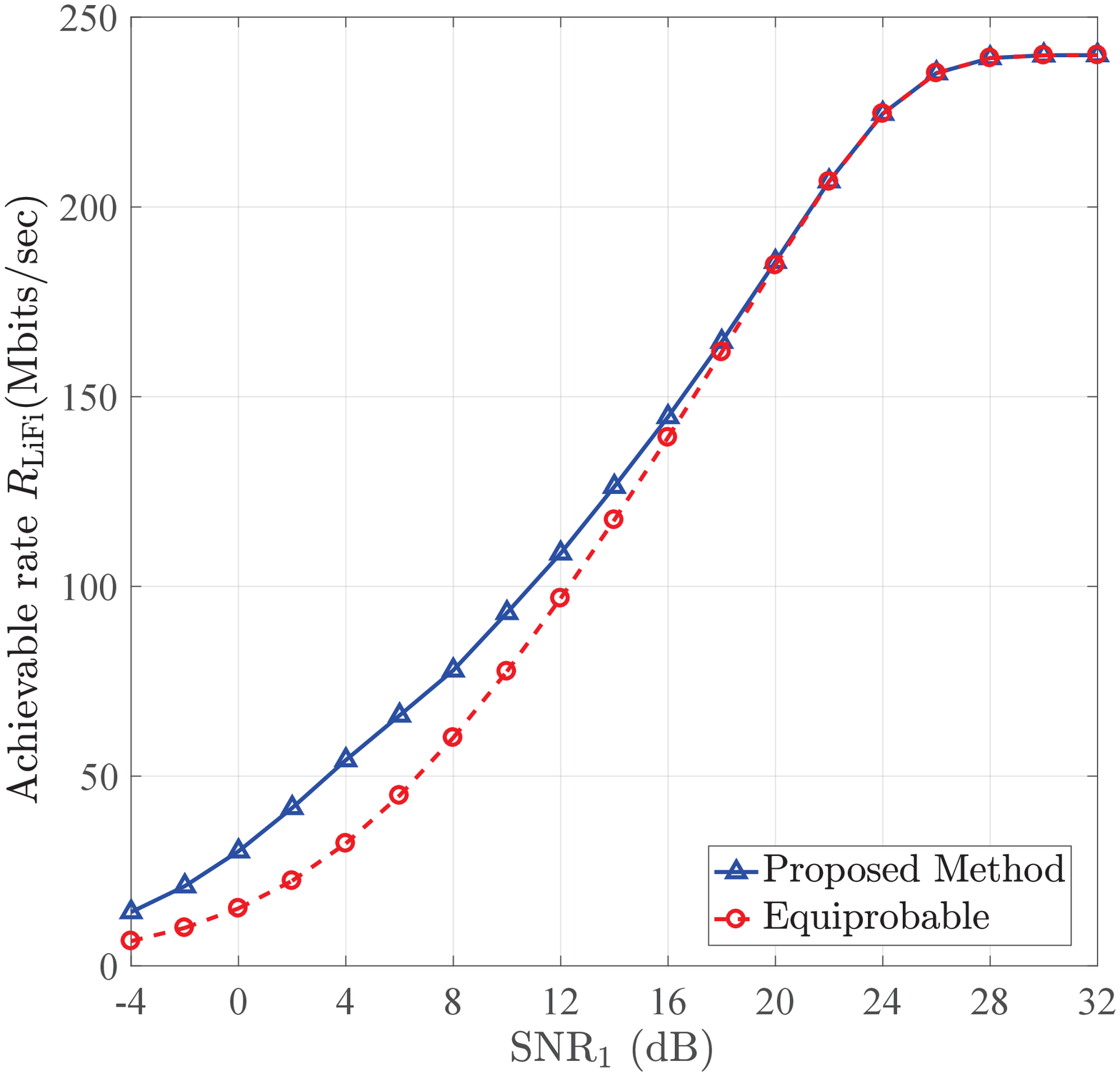}
        \vskip-0.2cm\centering {\footnotesize (c)}
    \end{minipage}\hfill

    \begin{minipage}[b]{0.45\textwidth}
        \centering
        \includegraphics[height=5cm]{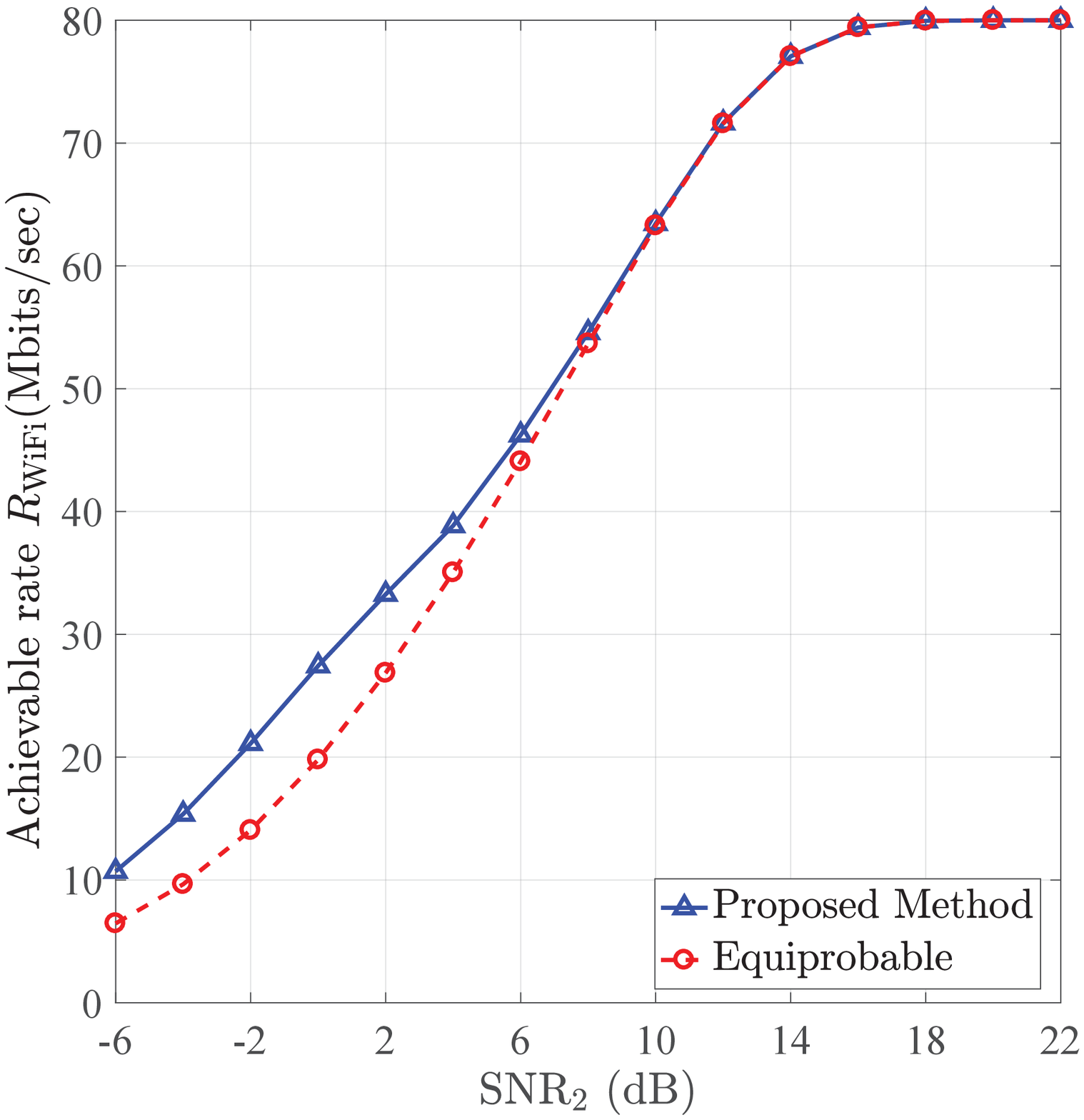}
        \vskip-0.2cm\centering {\footnotesize (d)}
    \end{minipage}\hfill
    \caption{(a)~The optimal probability distribution of input $\left\{ {{x_{1,k}}}, {{p_{1,k}}} \right\}$ versus ${\rm{SNR}}_1$ of LiFi link;(b)~The optimal probability distribution of input $\left\{ {{x_{2,l}}}, {{p_{2,l}}} \right\}$ with ${\rm{SNR}}_2$=4dB of WiFi link.
        (c)~Achievable rate of LiFi link ${{{R}}_{{\rm{LiFi}}}}$  versus ${\rm{SNR}}_1$;
        (d)~Achievable rate of WiFi link ${{{R}}_{{\rm{WiFi}}}}$  versus ${\rm{SNR}}_2$.
    }\label{WiFi_pro}
    \label{LiFi_pro}
\end{figure}

\begin{figure}[htbp]
    \begin{minipage}[b]{0.45\textwidth}
        \centering
        \includegraphics[height=6.5cm]{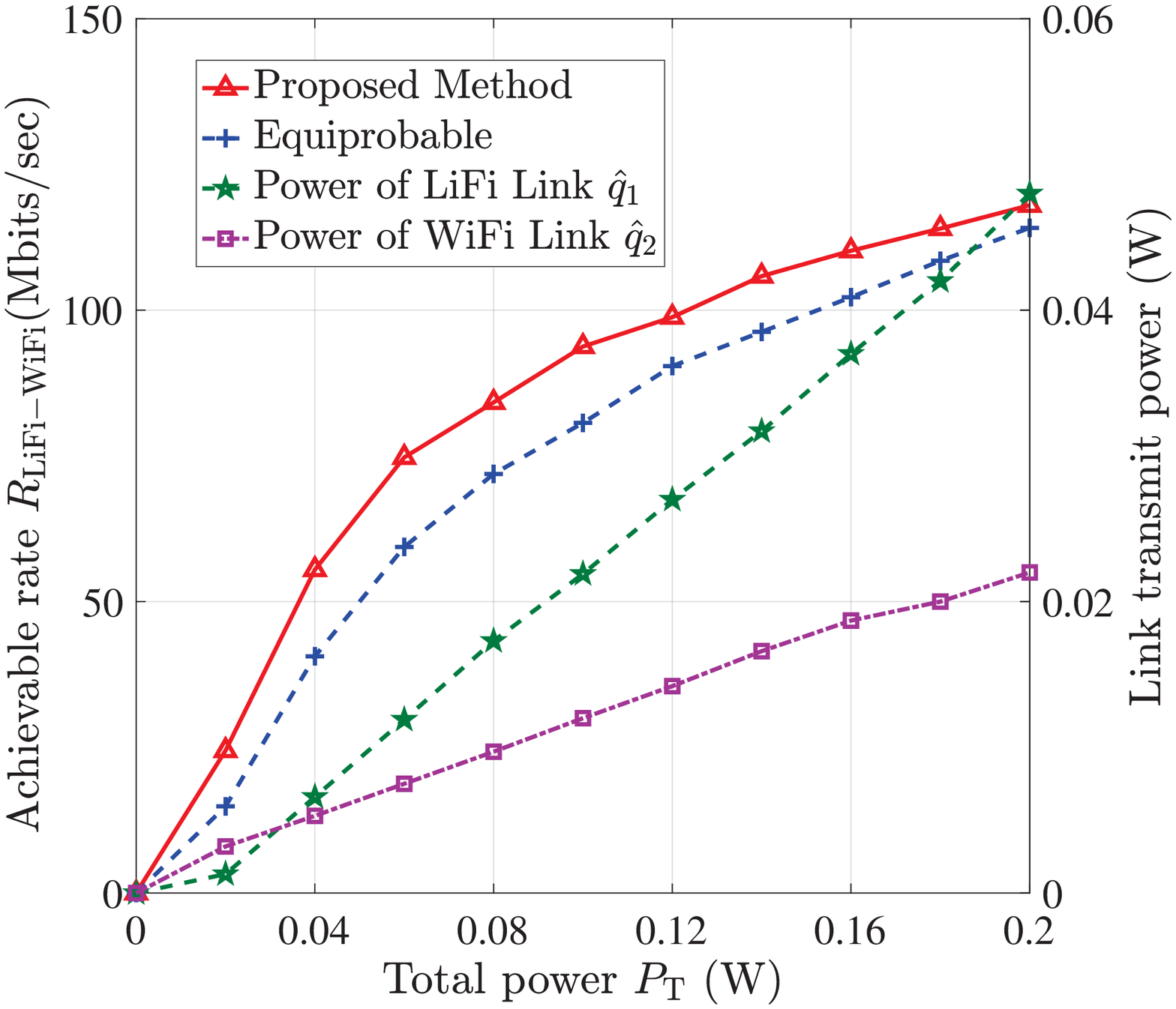}
        \vskip-0.2cm\centering {\footnotesize (a)}
    \end{minipage}\hfill
    \begin{minipage}[b]{0.45\textwidth}
        \centering
        \includegraphics[height=6.5cm]{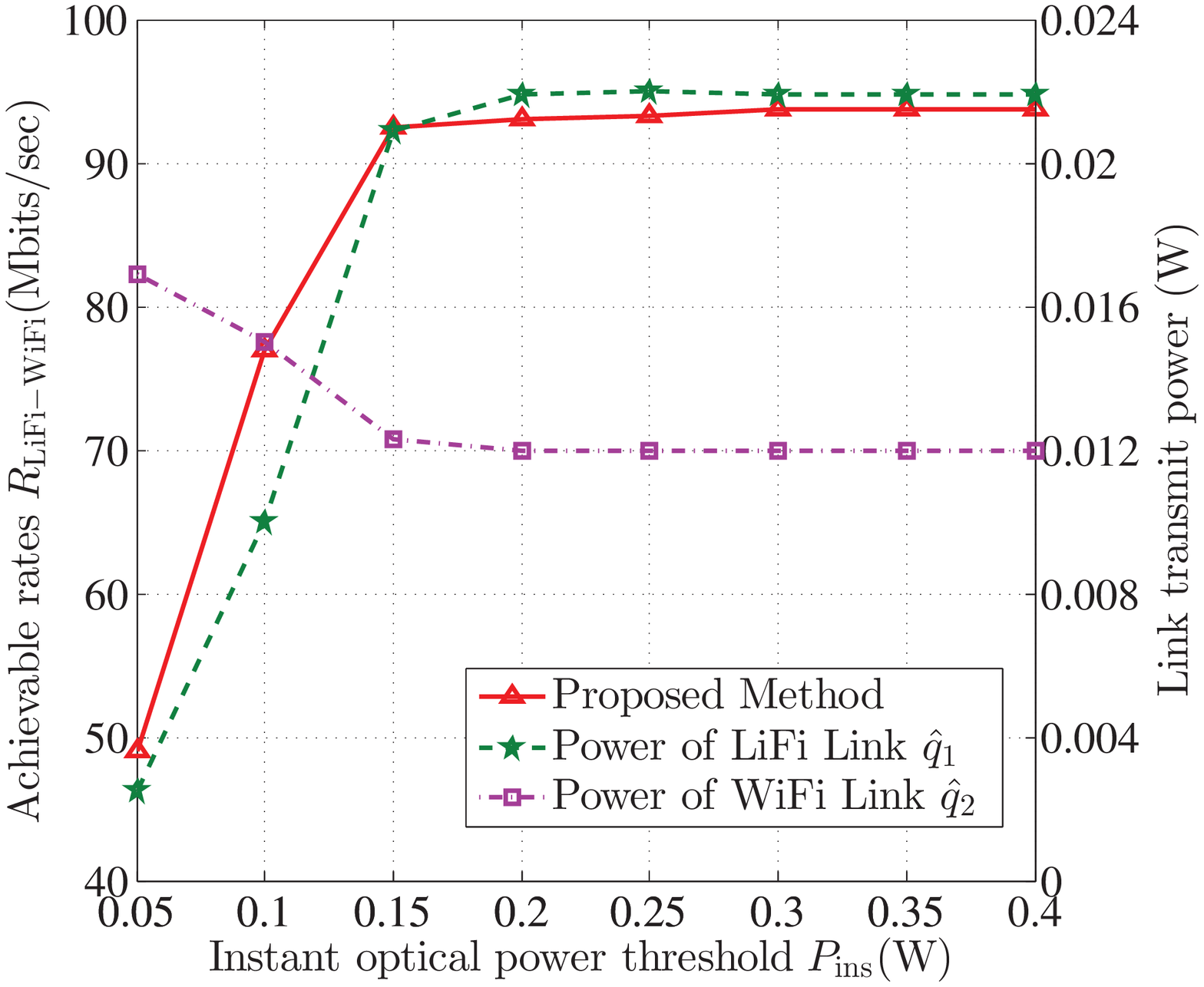}
        \vskip-0.2cm\centering {\footnotesize (b)}
    \end{minipage}\hfill
    \caption{Achievable rate of the aggregated
        LiFi-WiFi  system ${{{R}}_{{\rm{LiFi-WiFi}}}}$:  (a)~versus the total   power
        threshold $P_{\rm{T}}$;
        (b)~versus the instant  optical power threshold $P_{\rm{ins}}$.
    }\label{achievable}
\end{figure}

In the following, we evaluate the performance of the proposed transmission schemes in Fig.~\ref{LiFi_pro} and Fig.~\ref{achievable}.

Fig.~\ref{LiFi_pro} (a)  illustrates the optimal probability distribution of input $\left\{ {{x_{1,k}}}, {{p_{1,k}}} \right\}$
versus ${\rm{SNR}}_1$ (SNR) of the LiFi link,  where $K_1=8$. It shows that in the low SNR region, the optimal   input positions include two discrete points with an equal probability. While in the high SNR region, the optimal   input positions have more than two discrete points. As the increase of SNR, the optimal  probability distribution is closer to the equiprobable distribution.
Fig.~\ref{LiFi_pro} (c)  illustrates the achievable rate of LiFi link ${{{R}}_{{\rm{LiFi}}}}$  versus SNR. We can observe that the achievable rate of LiFi link ${{{R}}_{{\rm{LiFi}}}}$ with the proposed method is higher than that equiprobable   distribution.
Moreover, with the increase of  the SNR, the gap between the proposed method and the equiprobable   distribution   shrinks.

Fig.~\ref{WiFi_pro} (b)   shows the optimal probability distribution of input $\left\{ {{x_{2,l}}}, {{p_{2,l}}} \right\}$  with
  ${\rm{SNR}}_2=$ 4dB  of the WiFi link,  where $K_2=16$. It shows that for $\mathrm{SNR}_2$=4dB,  the optimal   input positions include sixteen discrete points, and  the optimal  probability distribution is  equiprobable.
Fig.~\ref{WiFi_pro}(d)  illustrates the achievable rate of WiFi link ${{{R}}_{{\rm{WiFi}}}}$  versus $\mathrm{SNR}_2$. We can observe that the achievable rate of WiFi link ${{{R}}_{{\rm{WiFi}}}}$ with the proposed method is higher than that obtained by the equiprobable   distribution.
Moreover, with the increase of  the $\mathrm{SNR}_2$, the gap between the proposed method and the equiprobable   distribution decreases.

Fig.~\ref{achievable} (a)  shows the achievable rate  ${{{R}}_{{\rm{LiFi-WiFi}}}}$ and link transmit power   of the aggregated
LiFi-WiFi  system   versus total   power
threshold $P_{\rm{T}}$, respectively.
We observe
that the achievable   rates  ${{{R}}_{{\rm{LiFi-WiFi}}}}$ with both the proposed method and equiprobable distribution increase as  total  power
threshold $P_{\rm{T}}$ increases.

Fig.~\ref{achievable} (b)  illustrates the achievable rate of aggregated
LiFi-WiFi  system  ${{{R}}_{{\rm{LiFi-WiFi}}}}$ versus instant optical power threshold $P_{\rm{ins}}$.
We observe
that the achievable   rates  ${{{R}}_{{\rm{LiFi-WiFi}}}}$ with the proposed method  increase as the instant  optical power threshold $P_{\rm{ins}}$ first increases, and remains constant.
Moreover, as instant  optical power threshold $P_{\rm{ins}}$ increases,  the  transmit power of LiFi link  $\hat q_1$   first increases,  and then
keeps as a constant.
The transmit power
of WiFi link $\hat q_2$  first decreases,  and then
keeps as a constant. This is due to $\tau \buildrel \Delta \over =\min \left( {{{{P_o}/\bar \mu}},{ {P_{\rm{ins}}}/A}{}} \right)$, and we assume that ${P_o}=0.8{P_{\rm{ins}}}$, and $\bar \mu  = 0.5 A$.
 For a lower $P_{\rm{ins}}$, $\hat q_1$ is constrained by $\tau$, while for a high $P_{\rm{ins}}$, $\hat q_1$ is constrained by the total electrical power
threshold $P_{\rm{T}}$.

\subsection{Optimal Discrete Constellation Input Distributions Based on   ${{{R}}^{\rm{L}}_{{\rm{LiFi-WiFi}}}}$  \eqref{rate_up_low}}

\begin{figure}[htbp]
    \begin{minipage}[b]{0.45\textwidth}
        \centering
        \includegraphics[height=6.5cm]{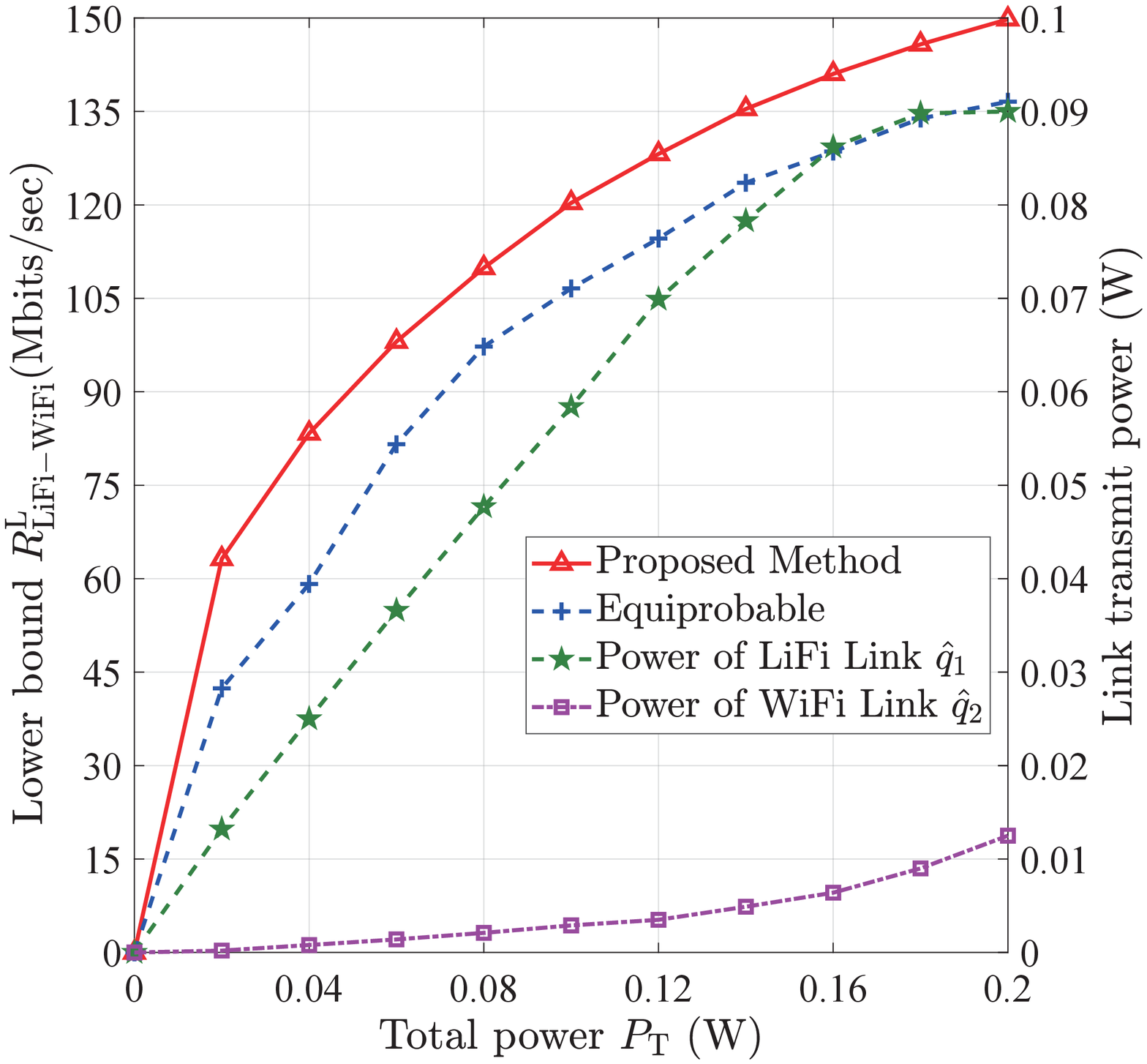}
        \vskip-0.2cm\centering {\footnotesize (a)}
    \end{minipage}\hfill
    \begin{minipage}[b]{0.45\textwidth}
        \centering
        \includegraphics[height=6.5cm]{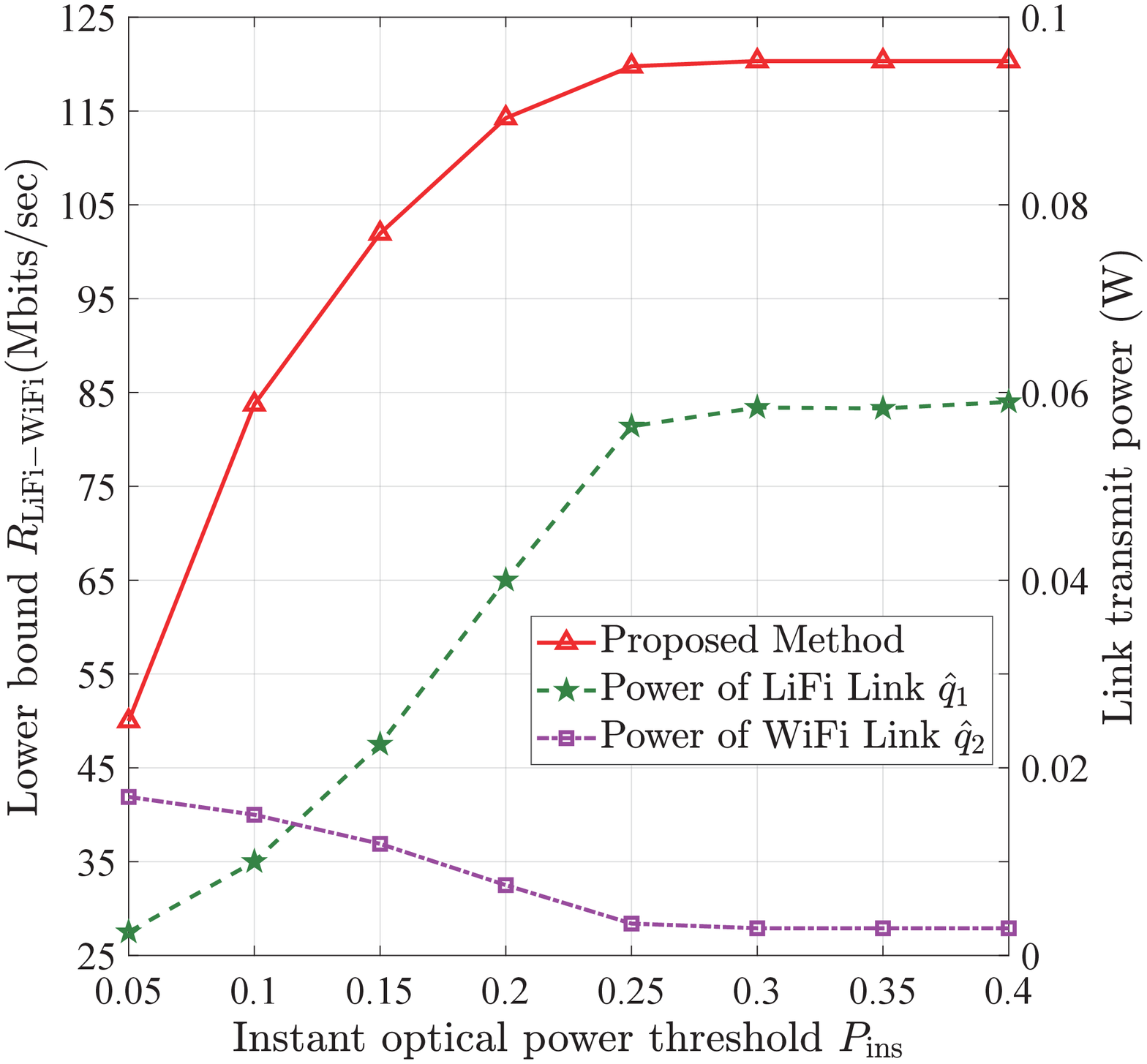}
        \vskip-0.2cm\centering {\footnotesize (b)}
    \end{minipage}\hfill
    \caption{Lower bound of the achievable rates of the aggregated
        LiFi-WiFi  system ${{{R}}^{\rm{L}}_{{\rm{LiFi-WiFi}}}}$:  (a)~versus the total electrical power
        threshold $P_{\rm{T}}$;
        (b)~versus the instant  optical power threshold $P_{\rm{ins}}$.
    }\label{lower_power}
\end{figure}

In the following, the performance of the proposed aggregated
LiFi-WiFi  system based on  the lower bound of   achievable rate is investigated in Fig.~\ref{lower_power}  (a) and (b).

Fig.~\ref{lower_power} (a)  demonstrates the lower bound of achievable rate  ${{{R}}^{\rm{L}}_{{\rm{LiFi-WiFi}}}}$ and link transmit power   of aggregated
LiFi-WiFi  system   versus total  power
threshold $P_{\rm{T}}$, respectively.
We observe
that the lower bound of  achievable   rates  ${{{R}}^{\rm{L}}_{{\rm{LiFi-WiFi}}}}$ with both the proposed method and equiprobable distribution increase as  total   power
threshold $P_{\rm{T}}$ increases.
Moreover, as total power threshold ${P_T}$ increases,    the transmit power of the LiFi link, i.e.,  $\hat q_1$,   first increases, and remains as a constant.and  the transmit power of the WiFi link, i.e., $\hat q_2$,  increases.The reason is that  $\hat q_1$  is also limited by the optical power constraint.\footnote{{Due to adding the constant gap, $R_{\mathrm{LiFi-WiFi}}$ might be less than $R_{\mathrm{LiFi-WiFi}}^{\mathrm{L}}$ in Fig.~\ref{lower_power}-\ref{compare_B} as similar as Fig. 1-3 of \cite{Zeng2012WCL}.}}

Fig.~\ref{lower_power} (b)  describes the  lower bound of achievable rate of aggregated LiFi-WiFi  system  ${{{R}}^{\rm{L}}_{{\rm{LiFi-WiFi}}}}$ versus instant optical power threshold $P_{\rm{ins}}$, and  we assume that ${P_o}=0.8{P_{\rm{ins}}}$, and $\bar \mu  = 0.5 A$. We observe that the  lower bound of achievable   rates  ${{{R}}^{\rm{L}}_{{\rm{LiFi-WiFi}}}}$ with the proposed method  increases as  instant  optical power  threshold $P_{\rm{ins}}$ first increases, and remains as a constant. Moreover, as instant  optical power threshold $P_{\rm{ins}}$ increases,   the transmit power of the LiFi link, i.e.,  $\hat q_1$,   first increases,  and then keeps as a constant. The transmit power of WiFi link $\hat q_2$  first decreases,  and then keeps as a constant. This is due to $\tau \buildrel \Delta \over =\min \left( {{{{P_o}/\bar \mu}},{ {P_{\rm{ins}}}/A}{}} \right)$.  For a lower $P_{\rm{ins}}$, $\hat q_1$ is constrained by $\tau$, while for high $P_{\rm{ins}}$, $\hat q_1$ is constrained by the total electrical power threshold $P_{\rm{T}}$.

\subsection{The Rate Performance}

\begin{figure}[htbp]
    \begin{minipage}[b]{0.45\textwidth}
        \centering
        \includegraphics[height=6.5cm]{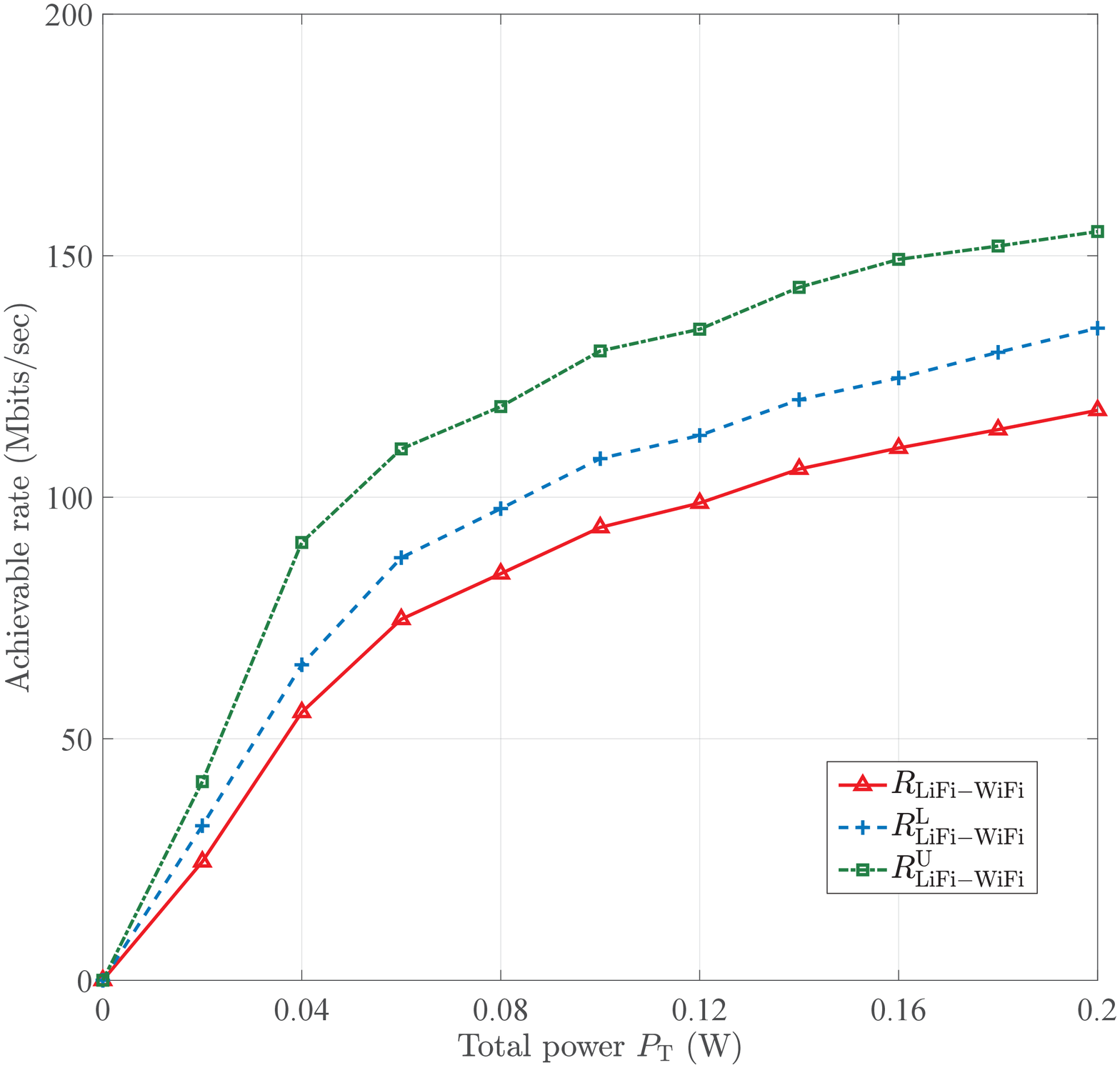}
        \vskip-0.2cm\centering {\footnotesize (a)}
    \end{minipage}\hfill
    \begin{minipage}[b]{0.45\textwidth}
        \centering
        \includegraphics[height=6.5cm]{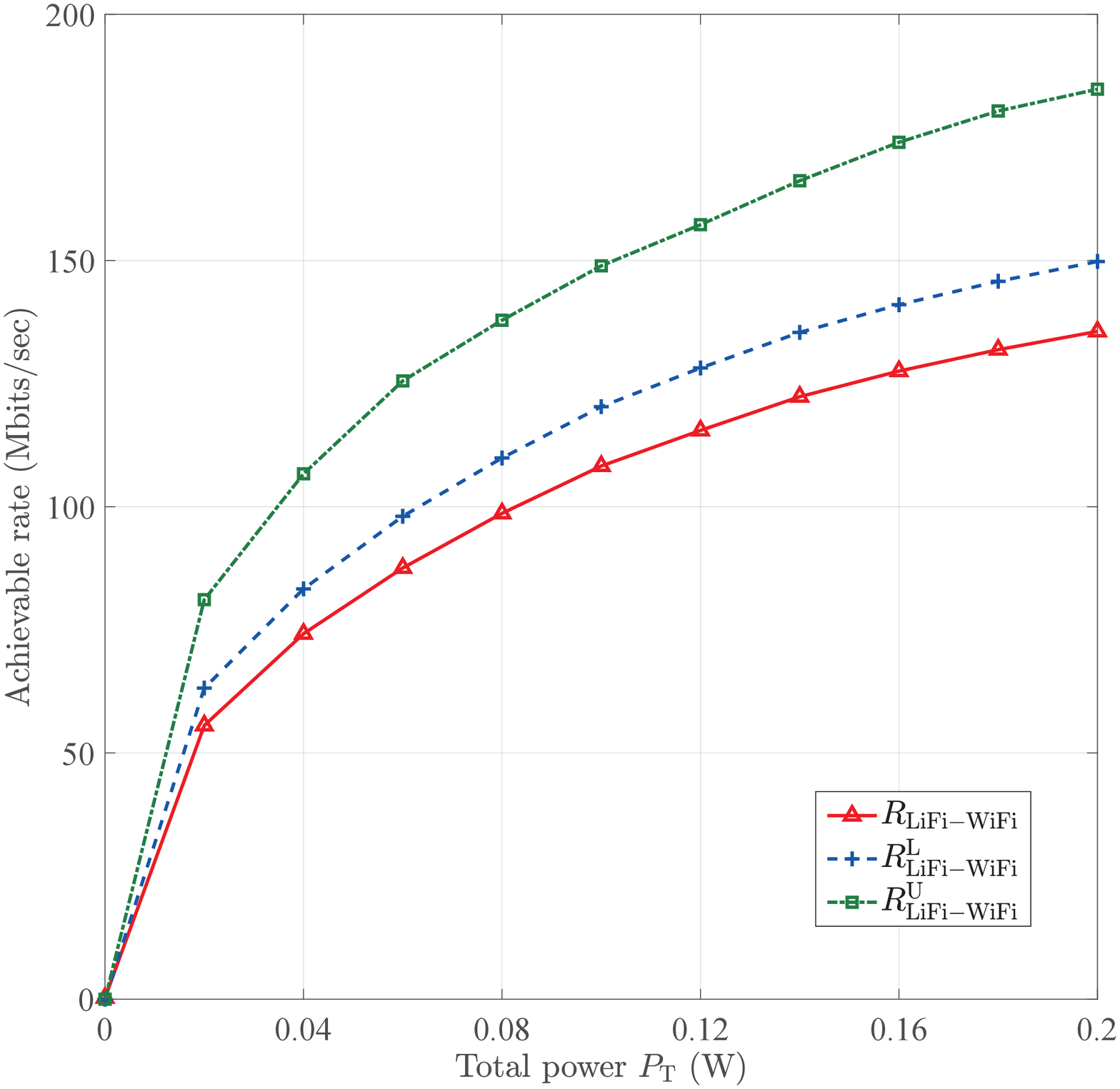}
        \vskip-0.2cm\centering {\footnotesize (b)}
    \end{minipage}\hfill
    \caption{${{{R}}_{{\rm{LiFi-WiFi}}}}$, ${{{R}}^{\rm{L}}_{{\rm{LiFi-WiFi}}}}$  and ${{{R}}^{\rm{U}}_{{\rm{LiFi-WiFi}}}}$   versus the total   power
        threshold $P_{\rm{T}}$: (a)~with the optimal solutions of problem \eqref{max_rate_accuracy};
        (b)~with the optimal solutions of problem \eqref{max_rate_agg}.
    }\label{compare}
\end{figure}

\begin{figure}[htbp]
    \begin{minipage}[b]{0.45\textwidth}
        \centering
        \includegraphics[height=6.5cm]{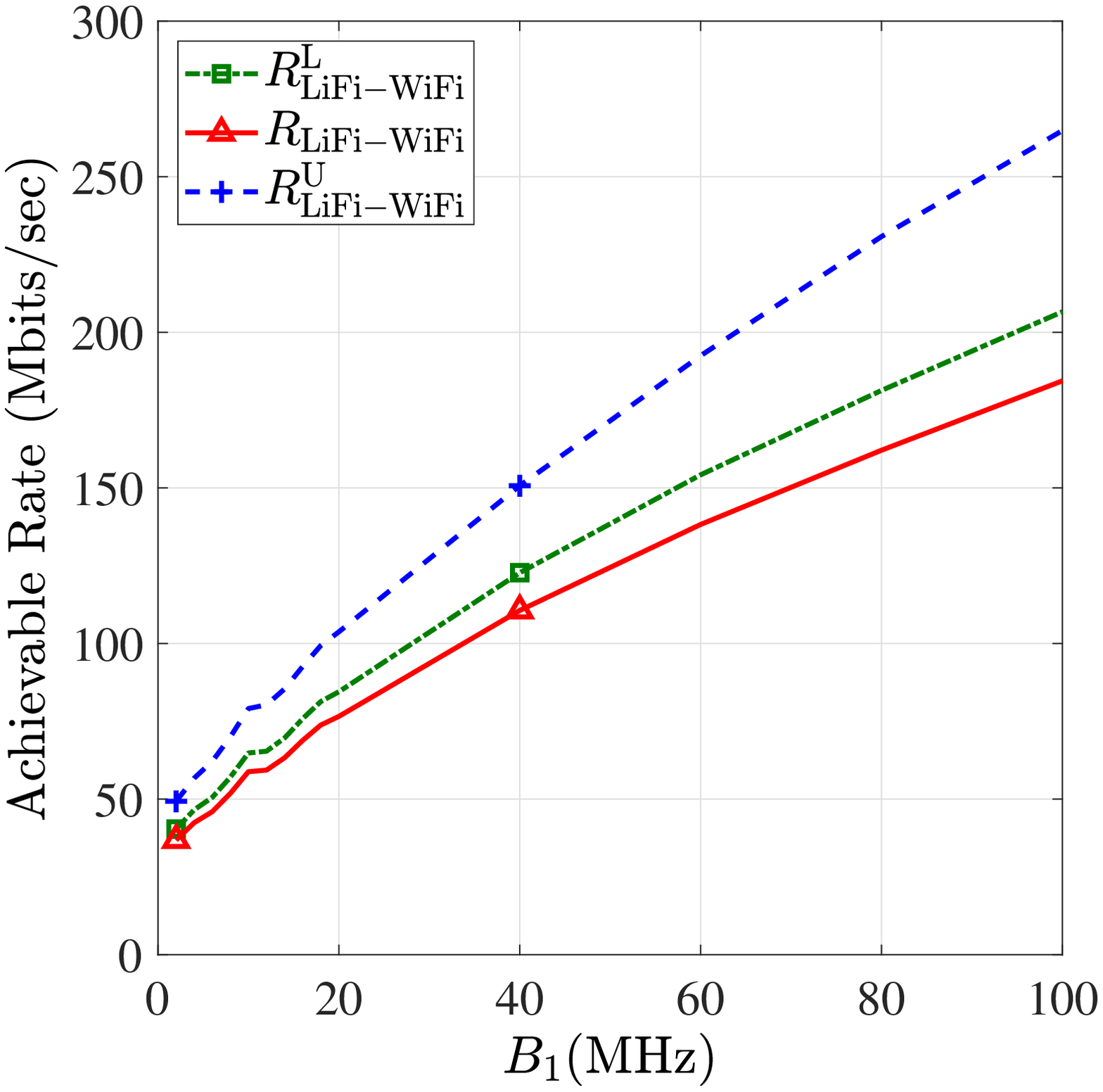}
        \vskip-0.2cm\centering {\footnotesize (a)}
    \end{minipage}\hfill
    \begin{minipage}[b]{0.45\textwidth}
        \centering
        \includegraphics[height=6.5cm]{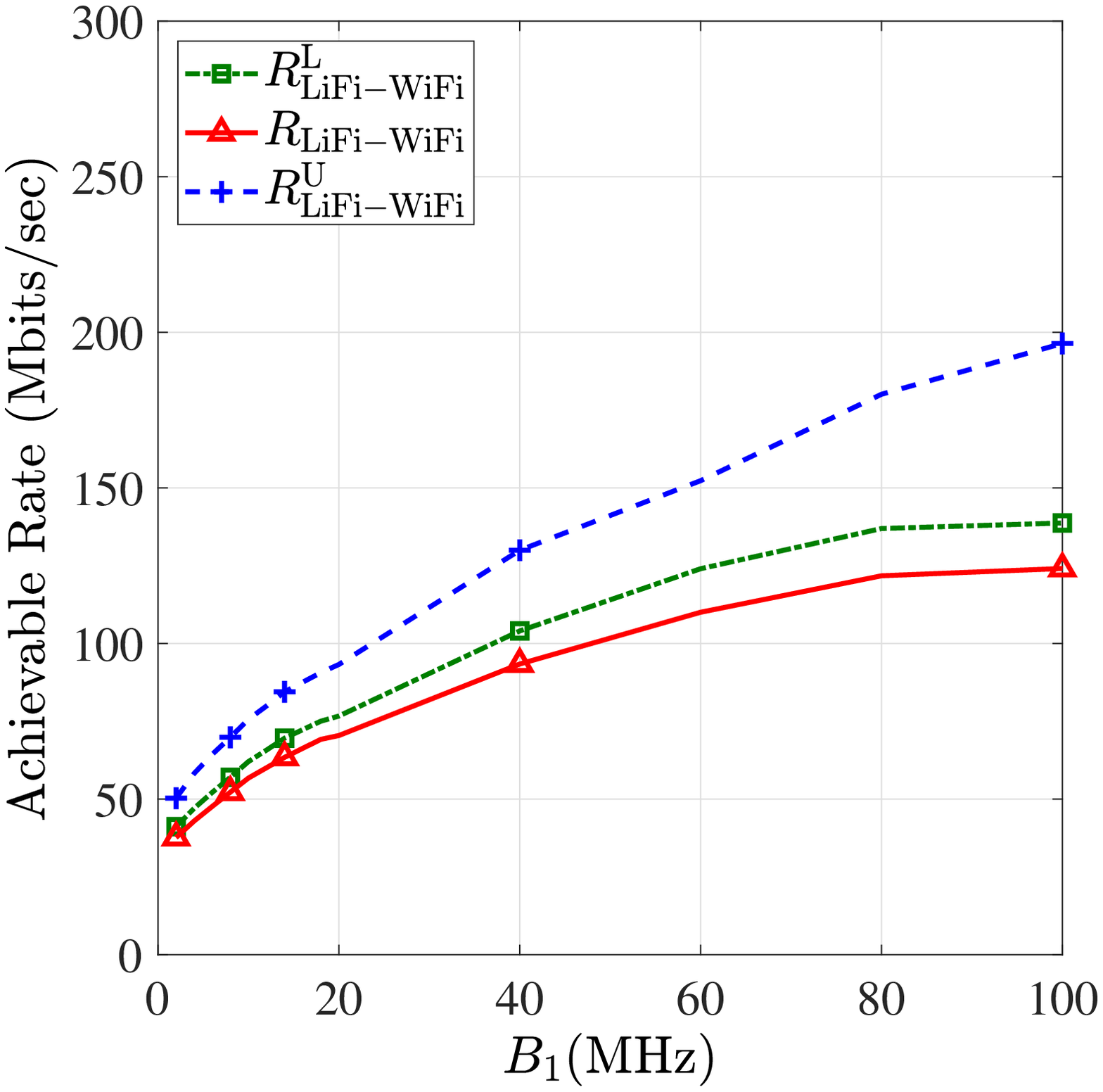}
        \vskip-0.2cm\centering {\footnotesize (b)}
    \end{minipage}\hfill
    \caption{${{{R}}_{{\rm{LiFi-WiFi}}}}$, ${{{R}}^{\rm{L}}_{{\rm{LiFi-WiFi}}}}$  and ${{{R}}^{\rm{U}}_{{\rm{LiFi-WiFi}}}}$   versus the LiFi link bandwidth $B_1$ with the WiFi link bandwidth $B_2 = 20\mathrm{MHz}$: (a)~with the optimal solutions of problem \eqref{max_rate_accuracy};
        (b)~with  the optimal solutions of problem \eqref{max_rate_agg}.
    }\label{compare_B}
\end{figure}

In the following, we investigate the  comparison between  the achievable rate ${{{R}}_{{\rm{LiFi-WiFi}}}}$  \eqref{rate_LiFi_WiFi}, it's lower bound ${{{R}}^{\rm{L}}_{{\rm{LiFi-WiFi}}}}$  \eqref{rate_up_low} and the upper bound ${{{R}}^{\rm{U}}_{{\rm{LiFi-WiFi}}}}$  \eqref{rate_up_up}  of the proposed aggregated
LiFi-WiFi  system  in Fig.~\ref{compare} (a) and (b), Fig.~\ref{compare_B} (a) and (b).

Fig.~\ref{compare} (a)  shows the achievable rate ${{{R}}_{{\rm{LiFi-WiFi}}}}$  \eqref{rate_LiFi_WiFi},  it's lower bound ${{{R}}^{\rm{L}}_{{\rm{LiFi-WiFi}}}}$  \eqref{rate_up_low} and the upper bound ${{{R}}^{\rm{U}}_{{\rm{LiFi-WiFi}}}}$  \eqref{rate_up_up} of the aggregated LiFi-WiFi  system   versus total   power threshold $P_{\rm{T}}$ with the optimal solutions of problem \eqref{max_rate_accuracy}, respectively. And Fig.~\ref{compare} (b)  illustrates the achievable rate ${{{R}}_{{\rm{LiFi-WiFi}}}}$  \eqref{rate_LiFi_WiFi}, it's lower bound ${{{R}}^{\rm{L}}_{{\rm{LiFi-WiFi}}}}$  \eqref{rate_up_low} and the upper bound ${{{R}}^{\rm{U}}_{{\rm{LiFi-WiFi}}}}$  \eqref{rate_up_up} of the aggregated LiFi-WiFi  system   versus total   power threshold $P_{\rm{T}}$ with the optimal solutions of problem \eqref{max_rate_agg}, respectively. Both Fig.~\ref{compare} (a) and (b) show that  for a low ${P_T}$, the gap between ${{{R}}_{{\rm{LiFi-WiFi}}}}$ and  ${{{R}}^{\rm{L}}_{{\rm{LiFi-WiFi}}}}$ is larger than the one of between ${{{R}}_{{\rm{LiFi-WiFi}}}}$ and ${{{R}}^{\rm{U}}_{{\rm{LiFi-WiFi}}}}$, and for a high ${P_T}$, the gap between ${{{R}}_{{\rm{LiFi-WiFi}}}}$ and ${{{R}}^{\rm{L}}_{{\rm{LiFi-WiFi}}}}$  is lower than the one of between ${{{R}}_{{\rm{LiFi-WiFi}}}}$ and ${{{R}}^{\rm{U}}_{{\rm{LiFi-WiFi}}}}$.
Moreover, the   ${{{R}}_{{\rm{LiFi-WiFi}}}}$ , ${{{R}}^{\rm{L}}_{{\rm{LiFi-WiFi}}}}$ and  ${{{R}}^{\rm{U}}_{{\rm{LiFi-WiFi}}}}$ in Fig.~\ref{compare} (a)  are lower than those in Fig.~\ref{compare} (b). This is because that solutions of Algorithm 3 are suboptimal.

{
        Fig.~\ref{compare_B} (a) and (b) shows that the achievable rate versus the LiFi link bandwidth $B_1$ with the WiFi link bandwidth $B_2 = 20\mathrm{MHz}$ and the two solutions. Fig. ~\ref{compare} (a) based on the solutions of problem \eqref{max_rate_accuracy}, and Fig. ~\ref{compare} (b) is with the solutions of problem \eqref{max_rate_agg}.
        Both Fig.~\ref{compare_B} (a) and (b) show that  for a low ${B_1}$, the gap between ${{{R}}_{{\rm{LiFi-WiFi}}}}$ and  ${{{R}}^{\rm{L}}_{{\rm{LiFi-WiFi}}}}$ is larger than the one of between ${{{R}}_{{\rm{LiFi-WiFi}}}}$ and ${{{R}}^{\rm{U}}_{{\rm{LiFi-WiFi}}}}$, and for a high ${B_1}$, the gap between ${{{R}}_{{\rm{LiFi-WiFi}}}}$ and ${{{R}}^{\rm{L}}_{{\rm{LiFi-WiFi}}}}$  is lower than the one of between ${{{R}}_{{\rm{LiFi-WiFi}}}}$ and ${{{R}}^{\rm{U}}_{{\rm{LiFi-WiFi}}}}$.
}

{ \subsection{The Average Executing Time}
    To evaluate the computational complexity, Table II shows the average executing times versus modulation order ($M = N$)  of Algorithm \ref{alg:Framwork_O_2} and Algorithm \ref{alg:Framwork_L_2}, respectively.
    
    \begin{table}[htbp]
        \centering
        \caption{The average executing time of different scheme versus modulation order($M = N$).}\label{ave_exe_time}
        \begin{tabular}{|c|c|c|c|c|c|}
            \hline
            \diagbox{$M=N$} {Time(s)} {Scheme} & Algorithm \ref{alg:Framwork_O_2} &Algorithm \ref{alg:Framwork_L_2}\\
            \hline
             $4$  &$5.47$    &$0.2466$\\
             \hline
             $8$  &$30.16$   &$4.60$\\
             \hline
             $16$ &$79.01$   &$7.68$\\
             \hline
             $32$ &$311.44$  &$29.75$\\
             \hline
             $64$ &$4217.26$ &$264.95$\\
            \hline
        \end{tabular}
    \end{table}
    
    For the given modulation order, the average executing time is calculated over $200$ independent trials, which are performed by MATLAB (2021a) with Intel(R) Core(TM) i5-9500 3.00 GHz CPU and 8 GB RAM.
    As showed in Table \ref{ave_exe_time},   the CPU time of the two  algorithms increases as the modulation order increases, and the CPU time of Algorithm \ref{alg:Framwork_L_2} is significantly less than that of Algorithm \ref{alg:Framwork_O_2}.
}

\section{Conclusion}

  In this paper,  we investigated the    aggregated LiFi-WiFi system, which simultaneously used both
 the LiFi link and WiFi link to transmit information. First of all, with both
 LiFi link and WiFi link bandwidths consideration, we derived the achievable rate expression of the system with the discrete constellation input signals. Then, we further investigated the optimal input distribution and power allocation for the considered system. Moreover, we derived the upper and lower bounds of  the achievable rate expression of the system with the discrete constellation input signals, and we   investigated the optimal input distribution and power allocation for the considered system.
 At last, the effects of critical parameters of the aggregated
LiFi-WiFi  system on the maximum achievable rate, such as
total  threshold, optical power,   and
bandwidth, etc., are numerically analyzed.
In the future, we would further investigate both imperfect
CSI and interference management for multi-user LiFi-WiFi
networks.

  \begin{appendices} 
   \section{Proof of the formulation Lemma 1}\label{app_lemma_1}

The
achievable   rates of the LiFi link ${{{R}}_{{\rm{LiFi}}}}$     are given as follows

\begin{subequations}
\begin{align}
 {R_{{\rm{LiFi}}}} 
&  = {\rm{h}}\left( {{y_1}} \right) - {\rm{h}}\left( {{y_1}\left| {{x_1}} \right.} \right) \\
&=  - \int_{ - \infty }^\infty  {\frac{{\sum\limits_{k = 1}^M {{p_{1,k}}\exp \left( { - \frac{{z_{\rm{1}}^{\rm{2}}}}{{2\sigma _1^2}}} \right)} }}{{\sqrt {2\pi } {\sigma _1}}}{{\log }_2}} \frac{{\sum\limits_{m = 1}^M {{p_{1,m}}\exp \left( {{{\overline \Lambda  }_{k,m}}} \right)} }}{{\sqrt {2\pi } {\sigma _1}}}\,\mathrm{d}{z_1}\nonumber\\
    &~~~- \frac{1}{2}{\log _2}2\pi e\sigma _1^2 \\
& = -    \frac{1}{{2\ln 2}} - \sum\limits_{k = 1}^{{M}} {{p_{1,k}}{{\mathbb{E}}_{z_1}}\left\{ {{{\log }_2}\sum\limits_{m = 1}^{{M}} {{p_{1,m}}\exp \left( { {\overline \Lambda  _{k,m}}} \right)} } \right\}},
  \end{align}
\end{subequations}
where 
${\overline \Lambda  _{k,m}} \buildrel \Delta \over =  - \frac{{{{\left( {{g_1}{q_1}\left( {{x_{1,k}} - {x_{1,m}}} \right) + {z_1}} \right)}^{\rm{2}}}}}{{2\sigma _1^2}}$.

Suppose that  the bandwidth of the LiFi link is  $B_{ 1}$ (Hz). Then,
both the input and  output can be represented  by samples taken by $\frac{1}{{2B_{ 1}}}$ seconds apart \cite{Hranilovic}.
    The power spectral density of  the noise is  $\frac{\sigma_1 ^2}{2}$ watts/Hz, which leads to the noise   power
${{\sigma _1^2}}B_{ 1}$.
For the time interval $\left[ {0,T} \right]$, there are $2BT$ noise
samples, and the variance of each sample is  $\frac{{{\sigma _1^2}B_{ 1}T}}{{{2}B_{ 1}T}} = \frac{\sigma_1 ^2}{2}$.
Moreover, if the power of signal is $P_1$,   the
energy of signal per sample is $\frac{{{P_1}T}}{{{2}B_{ 1}T}} = \frac{P_1}{{{2}B_{ 1}}}$.
Therefore,  we obtain  the achievable rate  expression of   the LiFi link   with the bandwidth of  $B_1$ as     given in \eqref{rate_LiFi}.

Then, the
achievable   rates of the  WiFi link
     and ${{{R}}_{{\rm{WiFi}}}}$  are given as follows
\begin{subequations}
\begin{align}
{R_{{\rm{WiFi}}}}
&=  - \int_{ - \infty }^\infty  {\frac{{\sum\limits_{l = 1}^N {{p_{2,l}}\exp \left( { - \frac{{{{\left| {{z_2}} \right|}^{\rm{2}}}}}{{\sigma _{\rm{2}}^{\rm{2}}}}} \right)} }}{{\pi \sigma _{\rm{2}}^{\rm{2}}}}} {\log _2}\frac{{\sum\limits_{n = 1}^N {{p_{2,n}}\exp \left( {\overline \Gamma  _{l,n}} \right)} }}{{\pi \sigma _{\rm{2}}^{\rm{2}}}}\,\mathrm{d}{z_2}\nonumber\\
&~~~- {\log _2}\pi e\sigma _2^2 \\
&  =  -   \frac{{1}}{{\ln 2}} - \sum\limits_{l = 1}^{{N}} {{p_{2,l}}{{\mathbb{E}}_{{z_2}}}\left\{ {{{\log }_2}\sum\limits_{n = 1}^{{N}} {{p_{2,n}}\exp \left( {\overline \Gamma  _{l,n}} \right)} } \right\}},
  \end{align}
\end{subequations}
where ${\overline \Gamma  _{l,n}} \buildrel \Delta \over =  - \frac{{{{\left| {g_{\rm{2}}^{\rm{*}}{q_{\rm{2}}}\left( {{x_{2,l}} - {x_{2,n}}} \right) + {z_2}} \right|}^{\rm{2}}}}}{{\sigma _{\rm{2}}^{\rm{2}}}}$.

Suppose that  the bandwidth of the   WiFi  link is  $B_{ 2}$ (Hz). Then,
both the input and  output can be represented  by complex samples taken by $\frac{1}{{B_{ 2}}}$ seconds apart.
Note that since the noise is independent
in the I and Q components, each use of the complex channel can be thought of as two
independent uses of a real AWGN channel.  If the power of signal is $P_2$, the noise variance and the power constraint
per real symbol are $\frac{\sigma_2 ^2}{2} $ and $\frac{{{P_2}}}{{2{B_2}}}$,
 respectively.
Thus,  we have  the achievable rate  expression of   the WiFi link   with the bandwidth of  $B_1$ as     given in \eqref{rate_WiFi}.

Hence, the achievable rates  of   the aggregated LiFi-WiFi system  ${R_{{\rm{LiFi - WiFi}}}}$  is   given in (14).

\section{Derivation of the formulation Lemma 2}\label{app_lemma_2}

Since $\log_2 \left( {\sum {\exp \left( x \right)} } \right)$ is a convex function with respect to $x$,
we can obtain the  upper bound of  ${R_{{\rm{LiFi}}}}$  by Jensen's inequality.

\begin{subequations}\label{rate_LiFi_upper}
\begin{align}
 {R_{{\rm{LiFi}}}}  &\le  - \frac{1}{{2 \ln 2}} -  \sum\limits_{k = 1}^M {{p_{1,k}}} {\log _2}\sum\limits_{m = 1}^M {{p_{1,m}}\exp \left( { {{\mathbb{E}}_{z_1}} \left\{ {{{\overline \Lambda  }_{k,m}}} \right\}} \right)}  \\
 & = - \sum\limits_{k = 1}^M {{p_{1,k}}} {\log _2}\sum\limits_{m = 1}^M {{p_{1,m}}\exp \left( {2B_1\hat \Lambda_{k,m} - \frac{{{{\mathbb{E}}_{{z_{\rm{1}}}}}\left\{ {z_1^2} \right\}}}{{2\sigma _1^2}}} \right)}    \nonumber\\
 &~~~-\frac{1}{{2 \ln 2}} \\
 &=  - \sum\limits_{k = 1}^M {{p_{1,k}}} {\log _2}\sum\limits_{m = 1}^M {{p_{1,m}}\exp \left( 2B_1\hat \Lambda_{k,m} \right)}.
  \end{align}
\end{subequations}

The  upper bound    of    the LiFi link  ${R_{{\rm{LiFi}}}}$  with the bandwidth of  $B_1$  is     given by
\begin{align}
&{R_{{\rm{LiFi}}}}  \leq
 -2B_{1} \sum\limits_{k = 1}^{{M}} {{p_{1,k}}{{\log }_2}\sum\limits_{m = 1}^{{M}} {{p_{1,m}}} \exp \left( {2B_1\hat \Lambda_{k,m}} \right)}.
  \end{align}

Since $\log_2 \left( x \right)$ is a concave function,
we can obtain the  lower bound of  ${R_{{\rm{LiFi}}}}$ by Jensen's inequality as follows
\begin{subequations}\label{rate_LiFi_lowwer}
\begin{align}
 {R_{{\rm{LiFi}}}} & \ge  -   \frac{1}{{2\ln 2}} - \sum\limits_{k = 1}^{{M}} {{p_{1,k}}{{\log }_2}\sum\limits_{m = 1}^{{M}} {{p_{1,m}}} {{\mathbb{E}}_{z_1}}\left\{ {\exp \left( { {{{\overline \Lambda  }_{k,m}}}} \right)} \right\}} \\
 & =  -  \frac{1}{{2\ln 2}} + \frac{1}{{2}}- \sum\limits_{k = 1}^{{M}} {{p_{1,k}}{{\log }_2}\sum\limits_{m = 1}^{{M}} {{p_{1,m}}} \exp \left( {B_1\hat \Lambda_{k,m}} \right)}.
  \end{align}
\end{subequations}
The  lower bound  of    the LiFi link  ${R_{{\rm{LiFi}}}}$  with the bandwidth of  $B_1$  is     given by
\begin{align}
{R_{{\rm{LiFi}}}}  \geq
B_{1}- \frac{B_{1}}{{\ln 2}}    - 2B_{1}\sum\limits_{k = 1}^{{M}} {{p_{1,k}}{{\log }_2}\sum\limits_{m = 1}^{{M}} {{p_{1,m}}} \exp \left( { \hat \Lambda_{k,m}} \right)}.
  \end{align}

\section{Derivation of the formulation Lemma 3}\label{app_lemma_3}
The upper bound  of  ${R_{{\rm{WiFi}}}}$  is given as
\begin{subequations}
\begin{align}
 &{R_{{\rm{WiFi}}}}  \le  - \frac{1}{{\ln 2}} - \sum\limits_{l = 1}^N {{p_{2,l}}} {\log _2}\sum\limits_{n = 1}^N {{p_{2,n}}\exp \left( {  {{\mathbb{E}}_{{z_{\rm{2}}}}}\left\{ {{\overline \Gamma  _{l,n}}} \right\}} \right)}   \\
  & =   - \sum\limits_{l = 1}^N {{p_{2,l}}} {\log _2}\sum\limits_{n = 1}^N {p_{2,n}}\exp \left( 2{B_2}{{\hat \Gamma }_{l,n}} - {{\mathbb{E}}_{{z_{\rm{2}}}}}\left\{ {\frac{{{{\left| {{z_2}} \right|}^{\rm{2}}}}}{{\sigma _{\rm{2}}^{\rm{2}}}}} \right\} \right) \\
 & =  - \sum\limits_{l = 1}^N {{p_{2,l}}} {\log _2}\sum\limits_{n = 1}^N {{p_{2,n}}\exp \left( { 2B_2{\widehat\Gamma _{l,n}}} \right)}.
  \end{align}
\end{subequations}
Then, the  upper bound   of    the WiFi link  ${R_{{\rm{WiFi}}}}$  with the bandwidth of  $B_2$  is     given by
\begin{align}
 &{R_{{\rm{WiFi}}}} \leq  -  B_{2}\sum\limits_{l = 1}^{{N}} {{p_{2,l}}{{\log }_2}\sum\limits_{n = 1}^{{N}} {{p_{2,n}}} \exp \left( { 2{\widehat\Gamma _{l,n}}} \right)}.
  \end{align}

Since $\log_2 \left( x \right)$ is a concave function,
the lower bound of  ${R_{{\rm{WiFi}}}}$  can be obtained by Jensen's inequality as well
\begin{subequations}
\begin{align}
 {R_{{\rm{WiFi}}}} & \ge  - \frac{1}{{\ln 2}} - \sum\limits_{l = 1}^N {{p_{2,l}}{{\log }_2}\sum\limits_{n = 1}^N {{p_{1,n}}} \frac{1}{{\sigma _2^2\pi }}\int\limits_{ - \infty }^\infty  {\int\limits_{ - \infty }^\infty  {\exp \left( {} \right.} } }  \nonumber\\
 &~~\left. { - \frac{{{{\left( {{c_{\rm{R}}} + {z_{2,{\rm{R}}}}} \right)}^2} + z_{2,{\rm{R}}}^2}}{{{\sigma ^2}}} - \frac{{{{\left( {{c_{\rm{I}}} + {z_{2,{\rm{I}}}}} \right)}^2} + z_{2,{\rm{I}}}^2}}{{{\sigma ^2}}}} \right)d{z_{2,{\rm{R}}}}d{z_{2,{\rm{I}}}}  \label{lower_rf_b}  \\
   = & - \frac{1}{{\ln 2}} + 1 - \sum\limits_{l = 1}^{{N}} {{p_{2,l}}{{\log }_2}\sum\limits_{n = 1}^{{N}} {{p_{{2},n}}} \exp \left( { B_2{\widehat\Gamma _{l,n}}} \right)},
  \end{align}
\end{subequations}
where ${z_{2,{\rm{R}}}} \buildrel \Delta \over = {{\mathop{\rm Re}\nolimits} } \left( {z_{2}} \right)$,${z_{2,{\rm{I}}}}\buildrel \Delta \over ={{\mathop{\rm Im}\nolimits} } \left( {z_{2}} \right)$, ${c_{\rm{R}}}\buildrel \Delta \over ={{\mathop{\rm Re}\nolimits} } \left( {{{g}}^*_2{ q}_2\left( {{x_{2,l}} - {x_{2,n}}} \right)} \right) $ and ${c_{\rm{R}}}\buildrel \Delta \over ={{\mathop{\rm Im}\nolimits} } \left( {{{g}}^*_2{ q}_2\left( {{x_{2,l}} - {x_{2,n}}} \right)} \right) $;
 \eqref{lower_rf_b} is true due to ${{\mathbb{E}}_{z_2}}\left\{ {f\left( {z_2} \right)} \right\} = \int\limits_{z_2} {{p_{z_2}}} f\left( {z_2} \right){d_{z_2}}$, and $z_2$ follows the complex Gaussian distribution.

Finally, the   lower bound  of    the WiFi link  ${R_{{\rm{WiFi}}}}$  with the bandwidth of  $B_2$  is     given by
\begin{align}
{R_{{\rm{WiFi}}}} \geq B_{2}- \frac{B_{2}}{{\ln 2}}     - B_{2} \sum\limits_{l = 1}^{{N}} {{p_{2,l}}{{\log }_2}\sum\limits_{n = 1}^{{N}} {{p_{2,n}}} \exp \left( {{\widehat\Gamma _{l,n}}} \right)}\label{rate_WiFi_L1}.
  \end{align}

  \end{appendices}

\bibliographystyle{IEEE-unsorted}
\bibliography{refs0412}

\end{document}